%% file: ms.tex
\documentclass[journal, a4paper]{article}

\usepackage{amsmath, epsfig, times, amsthm, amsfonts, amssymb, comment, subfigure, multirow, algorithm, algorithmic}

\begin{document}

\title{Simple Countermeasures to Mitigate the Effect of Pollution Attack in Network Coding Based Peer-to-Peer Live Streaming}

\author{
	Attilio~Fiandrotti$^1$,
	Rossano~Gaeta$^2$,
	and~Marco~Grangetto$^2$
}
\date{%
    $^1$attilio.fiandrotti@polito.it\\%
    $^2$rossano.gaeta@unito.it, marco.grangetto@unito.it\\[2ex]%
    \today
}

\maketitle

\begin{abstract}

Network coding based peer-to-peer streaming represents an effective solution to aggregate user capacities and to increase system throughput in live multimedia streaming. Nonetheless, such systems are vulnerable to pollution attacks where a handful of malicious peers can disrupt the communication by transmitting just a few bogus packets which are then recombined and relayed by unaware honest nodes, further spreading the pollution over the network. Whereas previous research focused on malicious nodes identification schemes and pollution-resilient coding, in this paper we show pollution countermeasures which make a standard network coding scheme resilient to pollution attacks. Thanks to a simple yet effective analytical model of a reference node collecting packets by malicious and honest neighbors, we demonstrate that i) packets received earlier are less likely to be polluted and ii) short generations increase the likelihood to recover a clean generation. Therefore, we propose a recombination scheme where nodes draw packets to be recombined according to their age in the input queue, paired with a decoding scheme able to detect the reception of polluted packets early in the decoding process and short generations. The effectiveness of our approach is experimentally evaluated in a real system we developed and deployed on hundreds to thousands peers. Experimental evidence shows that, thanks to our simple countermeasures, the effect of a pollution attack is almost canceled and the video quality experienced by the peers is comparable to pre-attack levels.

\end{abstract}

\input{introduction}

\input{background}
\input{architecture}

%
\input{torostream}

\input{experiments}

\input{stateoftheart}
\input{conclusions}
\bibliographystyle{plain}
\bibliography{refs}

\end{document}

%% file: introduction.tex
\section{Introduction}
\label{sec:intro}

Peer-to-peer (P2P) video streaming represents a mature area of research with several successful examples to date \cite{zhang2005coolstreaming,huang2007pplive}.
The combination of P2P and Network Coding (NC) has recently received a great deal of attention from the research community as an effective mechanism to aggregate user capacities and to increase system throughput~\cite{p2p-nc-1,p2p-nc-2,p2p-nc-3}.
In NC-based architectures, the content is organized in independently decodable data units ({\em chunks} or {\em generations}) and each chunk is further partitioned in $k$ blocks.
The network nodes create linear combinations of suck blocks and produce coded packets that are transmitted to the network. The packets can be spread in the overlay network using a {\em push} approach, where at each transmission opportunity a new coded packet is generated by a peer and forwarded to a neighbor. On the receiver side, a chunk can be decoded as soon as enough coded packets have been collected by solving the system of linear equations corresponding to the collected packets.

Nonetheless, network coding systems are affected by a major Achille's heel: they are vulnerable to attacks carried out by nodes that spread bogus data over the network with teh goal of disrupting the communication. These actions are commonly known as \textit{pollution attacks} \cite{ross_07,ross_05} and the attackers are termed as malicious nodes.

Several issues need to be addressed to design effective solutions to pollution attack and the most part of the approaches proposed in the literature propose a two-steps approach.
First, some pollution detection mechanism is introduced to allow honest peers to detect an ongoing pollution attack and, if possible, the source thereof.
Second, a proper reaction (e.g., blacklisting) is undertaken after the presence or the source of the attack has been identified \cite{MIS,Li_Lui,Jin,gaeta2014dip,gaeta2013identification}. Both pollution detection and in particular malicious nodes identification can be very complex tasks involving high computational and/or communication overhead.

\subsection*{Our contribution}

The key goal of this work is to exploit the degrees of freedom available in standard random NC to design a media streaming architecture that is inherently resilient to pollution attacks.
By comparison, most of the related literature focuses either on identification and isolation of the malicious nodes or on designing ad-hoc data verification techniques as discussed in Section~\ref{sec:related}.
To this end, the contributions of this work are manifold:

\begin{itemize}

\item The main contribution is a novel packet recombination strategy where the nodes draw the packets to recombine among those in the input buffers with a probability that grows with the age of the packet in the buffer.
Our recombination scheme dramatically reduces the probability that an honest node transmits a polluted packet, which is further lowered by dividing the media stream in short generations. 
By comparison, in traditional NC every packets are drawn for recombination with identical probability and the media stream is subdivided in long generation to maximize the code efficiency.
To put up with the somewhat lower code efficiency of our recombination policy, we propose a simple heuristic which restores the code efficiency to almost pre-attack levels and improves the overall network utilization efficiency.

\item Our findings are supported by an analytic model which enables to understand how pollution propagates in a random NC push-based P2P system as a function of parameters such as generation size and time.
Namely, we show that the probability that a node forwards a polluted packet to downstream peers is not constant, rather it \emph{grows with time}, which justifies out age-based packet drawing policy.
Also, we show that the probability that a node recovers a clean generation depends on the generation size, i.e. short generations are more likely to enable successful generation recovery.
While our model relies on some simplifying assumptions, yet it represents an adequate solution to qualitatively describe the packet collection activity of a reference peer whose packets providers can be either malicious or honest.

\item Next, we present a probabilistic pollution detection mechanism which enables a node to autonomously detect the presence of polluted packets in its input buffer even if the node has not yet recovered the generation and without the need of external keys or hashing functions.
We experimentally show that our pollution detection scheme enables a node to detect pollution attacks earlier than a deterministic scheme which relies on an external verification server, further throttling the propagation of pollution through the network.

\item Finally, the performance of our resilient-by-design pollution avoidance scheme is throughly evaluated on a real, full-fledged, NC-based P2P video streaming protocol~\cite{eusipco2012,bandcodestmm} by streaming a live video sequence to one thousand peers.
Thanks to our realistic testbed, we are able to assess not only the reduction in the propagated pollution entailed by our strategy, but also the effect thereof on the video quality as perceived by the user in terms of continuity index, i.e., the fraction of video frames correctly recovered.

\end{itemize}

The rest of this paper is organized as follows: in Section \ref{sec:background} we overview the basics of multicast video distribution with binary random network coding (NC); next in Section \ref{sec:model-all} we illustrate a simple pollution attack model and we analytically study the propagation of the polluted packets through the network due to the recombinations at the nodes, showing that packets received early by the nodes are less likely to be polluted and small generations increase the probability to recover clean generations at the nodes.
In Section \ref{sec:architecture} the techniques that we propose to combat pollution are presented, namely an algebraic detection mechanism based on Gaussian Elimination and a pollution resistant NC coding strategy that recombines with higher probability those packets that are less likely to be polluted.
In Section \ref{sec:protocol} we overview ToroStream, a push-based protocol for P2P video distribution via NC that we use for experimenting with our algorithms with thousands of nodes in the following Section \ref{sec:experiments}. 
The paper ends with  Section \ref{sec:conclusions} drawing our conclusions and future research.
Finally please note that, to easy the reader, we collect in  Tab.~\ref{tab:main_notation} all  the key notation used throughout the paper. 

\input{notation_table.tex}

%% file: notation_table.tex
\begin{table}[ht]
	\begin{center}
		\begin{tabular}{|c|l|}
		\hline
		\multicolumn{2}{|c|}{NC parameters} \tabularnewline \hline
		$k, k'$ & Generation size, num. pkts required to decode ($k'$ $\ge$ $k$) \tabularnewline 
		$x_i$ & $i$-th data block \tabularnewline
		$F_i = (y_i, g_i)$ & Coded packets (payload, encoding vector) \tabularnewline
		$c = (c_1, \dots c_R)$ & Recombination vector \tabularnewline
		$p_{r}$ & Prob. each packet in input buffer is drawn for recombination\tabularnewline
		$m_r$ & Minimum rank to start recombining  \tabularnewline
		$\epsilon_c,$ & Code overhead, $\epsilon_c = (k' -k) / k$ \tabularnewline
		\hline
		\multicolumn{2}{|c|}{Attack model} \tabularnewline \hline
		$N$; $N_h$,$N_m$ & Tot. num. of nodes; Num. of honest, malicious nodes \tabularnewline
		$p_{poll}$ & Pollution probability of malicious nodes \tabularnewline 
		$r_p$ & Number of polluted packets received ($r_p \le k'$)\tabularnewline
		$\epsilon_p,$ & Pollution overhead, $\epsilon_p = r_p / k$ \tabularnewline
		\hline  
		\multicolumn{2}{|c|}{Analytical mode parameters} \tabularnewline \hline
		$n$ & Number of uploaders to reference node \tabularnewline 
		$x$ & Number of malicious uploaders to reference node  \tabularnewline \hline
		\multicolumn{2}{|c|}{P2P and experimental settings} \tabularnewline \hline
		$B_v$ & Test video bitrate \tabularnewline
		$C_t$ & Generation duration \tabularnewline
		$t_b$ & Buffering time \tabularnewline
		$N_s$ & Maximum allowed neighborhood size \tabularnewline
		$B_s$, $B_p$ & Server, peer nodes bandwidth \tabularnewline
		
		\hline

		\end{tabular}
	\caption{Key notation used in the paper.}
	\label{tab:main_notation}
	\end{center}
\end{table}

%% file: background.tex
\section{Background}
\label{sec:background}

In this section we first overview a typical push-based NC scheme in an unstructured mesh network detailing the operations at the network nodes.
Next, we describe a sample pollution attack model based on the injection of bogus coded packets into the network and we exemplify the spreading of the pollution through the network nodes.

\subsection{Media Streaming with Network Coding}

The source node holds a media content which is to be distributed to a set of cooperating nodes which we assume are arranged into an unstructured, non-acyclic, mesh network and operate according to a random-push model.
The video is subdivided in chunks of data called \emph{generations} that are independently encoded and decoded at the network nodes
so to achieve finite playback delay.
Each generation $x$ is further subdivided into $k$ blocks of symbols $(x_1,...,x_{k})$ (simply \emph{``blocks"} in the following) of identical size, where $k$ is the \emph{generation size}. 
Whereas a typical video sequence is subdivided in a large number of generations, for the sake of simplicity in the following we assume that the video sequence is composed  by just one generation.
Periodically, each node in the network including the source is given a \emph{transmission opportunity}: i.e., it is allowed to transmit one packet to the network.
Initially, only the source owns the original video content and distributes it to the other nodes transmitting encoded packets as follows.
Let vector $g_{i}=(g_{i,1}, ..., g_{i,k})$, $g_{i,j} \in GF(2)$ be the \emph{encoding vector} associated to the $i$-th coded packet, where $g_{i,j}$ is selected such that
$P\{g_{i,j} = 1\} = \frac{1}{2} ~~\forall i$. 
The source produces a random linear combination on the original blocks as $y_i=\sum_{j=1}^{k} g_{i,j} x_j$, where the sum operator represents the bit-wise XOR operator and $y_i$ is the $i$-th encoded payload.
The node forwards the encoded packet $F_i = (y_i, g_i)$, that contains the encoded payload $y_i$ along with the corresponding encoding vector $g_i$, to another node drawn at random in the network.
\\
The nodes of the network receive encoded packets, store them in an input buffer and transmit random linear combinations thereof as follows at every transmission opportunity.
Let us assume that a node has received $r$ packets $(F_1, ..., F_r)$: the node is allowed to transmits a linear combination of the payloads of the received packets; the $m$-th recombined packet is computed as $y_m^r = \sum_{j=1}^{r} c_{m,j} y_j$, where $c_{m,j} \in GF(2)$ and $P\{c_{m,j} = 1\} = p_r=\frac{1}{2}$, i.e. each received packet is recombined with equal probability. It turns out that the corresponding $m$-th encoding vectors is $g_m^r = \sum_{j=1}^{r} c_{m,j} g_j$.
The result of the recombination is novel packet $F_m^r (y_m^r, g_m^r)$ which is transmitted to the outgoing link of the node.
The recombinations at the nodes increase the likelihood that the transmitted packet is linearly independent from all the packets previously collected by the receiver, thus increasing the network goodput.
Each time a node receives a packet that is linearly independent from the previously received packets we say that the packet is \emph{innovative}.
We call the number of linearly independent packets received at any time by a node for the generation as the \emph{rank} of the generation at the node: once the rank is equal to $k$, we say that the generation has \emph{full rank}.
At this point, the node solves the system of linear equations corresponding to the received packets (\em e.g., \em via Gaussian elimination) and recovers the generation, i.e. the original video content.
\\
In practical NC applications, a receiver must however typically collect $k' > k$ packets because not all received packets are innovative due to the random combinations and forwarding.
The penalty $\epsilon = \frac{k'-k}{k}$ is usually termed as \emph{code overhead} and corresponds to the ratio of network bandwidth wasted transmitting non innovative, hence useless, packets .

\subsection{Pollution Attack Model}
\label{sec:attack-model}

\begin{figure}
  \begin{center}
	\includegraphics[width=0.90\columnwidth]{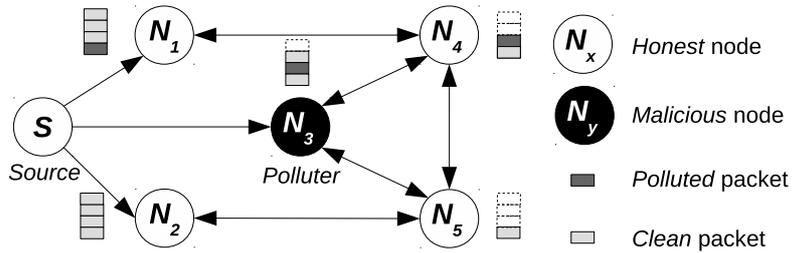}
  \end{center}
  \caption{Toy network with a source and $N$=5 nodes where $N_h$=4 are honest and $N_m$=1 ($N_3$) is malicious. The generation is composed of $k$=4 blocks and the nodes input buffers are represented at various decoding stages (polluted packets are represented in dark gray).}
  \label{fig:model-network}
\end{figure}

Let us assume that the overlay of network nodes is composed by one source node and $N$ peer nodes, where $N_h$ nodes are of the \emph{honest}
type and $N_m$ are of the \emph{malicious} type ($N_h + N_m = N$, where $N_m << N_h$) as depicted in Figure~\ref{fig:model-network}.
Honest nodes recombine the received packets as described in the previous section to allow as many other nodes as possible to recover the generation.
Malicious nodes disguise themselves among the honest ones and attempt to disrupt the video communication by randomly transmitting bogus coded data to the other network nodes.
At each transmission opportunity, the malicious node draws a random variable $\psi \in \{0, 1\}$ with uniform probability so that $\mathcal{P}\{\psi = 1\} = p_{poll}$.
If $\psi = 0$, the malicious node simply behaves as a honest one.
Otherwise if $\psi = 1$, the node generates a random encoding vector, a random encoded payload and transmits the packet to the network node: in this case, we say that the transmitted packet is \emph{polluted}.
Network nodes store the received packets in an input buffer without knowing if the packet is polluted or not, as shown in Figure~\ref{fig:model-network}.
Whenever a transmission opportunity arises for a honest node, if any of the $r_p$ polluted packets in its input buffer is drawn for recombination, then the transmitted packet is polluted too and bogus data is propagated to the other network nodes.
\\
In a scenario involving pollution attacks, we define as \emph{pollution overhead} the ratio $\epsilon_p = \frac{r_p}{k}$ of network bandwidth wasted transmitting packets that are polluted, hence useless.
Along with the previously defined code overhead, the pollution overhead will be used in this work to evaluate resources exploitation efficiency.
In the example of Figure \ref{fig:model-network}, node $N_3$ is malicious and has transmitted one polluted packet to $N_4$, which will not be able to correctly recover the generation.
Then, $N_4$ draws the polluted packet for recombination and transmits one packet to $N_1$: at this point also the input buffer of $N_1$ is polluted and the node will not be able to correctly recover the generation.

\section{Pollution Effects Model}
\label{sec:model-all}

In this section, we develop a simple analytical model to describe the behavior of a sample reference node that collects packets from a set of uploaders and combines them to forward a new packet to downstream nodes.
We show that the probability to correctly recover a generation increases with small generations, whereas the probability of forwarding a recombined polluted packet to downstream peers grows with time: these key observations are the basis to devise our proposed pollution-resilient packet recombination scheme proposed in Sect.~\ref{sec:proposed_architecture_recombination}.
Please note that we do not claim our model yields accurate predictions on the effect of pollution attacks on a real system. Indeed, the model is developed under several simplifying assumptions such as i) it describes the behavior of a randomly chosen (reference) peer in the overlay network; ii)
assumes that the overlay topology is an unstructured mesh where nodes all lay at the same hierarchical level iii) packets transmission happen at discrete time slots termed as a “rounds”; iv) during a round each uploader of the reference peer delivers a coded block. 
Nevertheless, the model includes all significant issues that determine the effect of polluting packets (and the effect of recombining polluted packets) before transmitting them to downstream peers as qualitatively (and, in part, quantitatively) experimentally verified later on.

\subsection{Modeling the Pollution Effects}
\label{sec:model}

\begin{figure}
  \begin{center}
	\includegraphics[width=0.85\columnwidth]{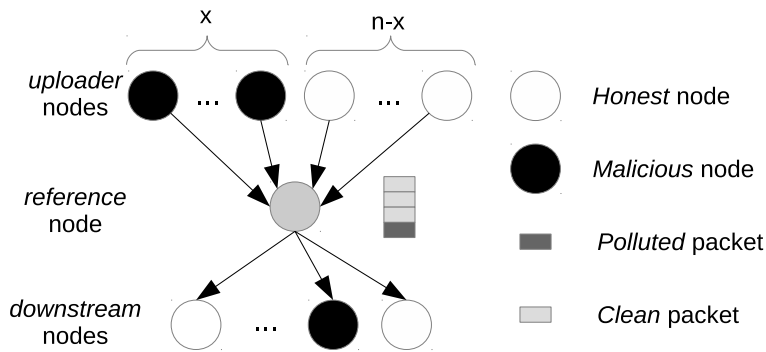}
  \end{center}
  \caption{Modeled scenario, where a reference node (middle of the picture, gray) receives packets from a set of uploaders, and transmits recombinations thereof to downstream nodes (malicious nodes are depicted in black).}
  \label{fig:model-scenario}
\end{figure}

%

To develop our model we consider a sample reference node that receives encoded packets from $n$ \emph{uploaders} nodes and forwards linear combinations thereof to other \emph{downstream} nodes as illustrated in Figure~\ref{fig:model-scenario} (the reference node is depicted in gray).
We assume that $x$ out of $n$ uploaders are malicious and purposely transmit bogus data as described in Sect.~\ref{sec:attack-model}.
To simplify the model derivation, we assume that time is discretized in {\em rounds}; during one round each of the $n$ uploaders delivers one packet to the reference node and the reference node transmits one packet to one of the downstream nodes.
\\
For the sake of simplicity, we assume that all packets received by a node are innovative and the number of rounds required to recover the generation is equal to $\lceil \frac{k}{n} \rceil + 1$.
The number of packets received by the reference node during the $i$-th round ($1 \leq i \leq \lceil \frac{k}{n} \rceil + 1$) is denoted as $R(i)$: under our assumptions $R(i)$ increases by $n$ at each round, hence $R(i)=i \cdot n$.

We denote as $P_p(i,x,b)$ the probability that $b$ out of the $n$ packets received at the $i$-th round are polluted when $x$ out of $n$ uploaders are malicious.
It is easy to show that this probability follows a binomial distribution, i.e.,
\begin{equation}
P_p(i,x,b) = \binom{ix}{b} p_{poll}^b (1-p_{poll})^{ix-b}.
\nonumber
\end{equation}
\noindent
Please note that since in one round each uploader delivers exactly one packet, the maximum number of polluted packets that our reference node can collect is equal to $i x$.

During the $i$-th round, the reference node draws at random a subset of the $R(i)$ packets contained in its input buffer and combines them to generate a new packet to forward to downstream nodes.
We compute the probability that the packet recombined by the reference node during the $i$-th round is polluted as
\begin{equation}
P_{rp}(i,x)=1-\sum_{b=0}^{ix} P_p(i,x,b) (1-p_r)^{b}.
\label{eq:recombined-polluted}
\end{equation}
that is, one minus the probability the recombined packet is not polluted (this probability is computed as the probability that none of the polluted packets received by the reference node has been selected for recombination).

We also assume the overlay network does not change with time and it is randomly built. Under these assumptions, we describe the probability that $x$ out of $n$ uploaders are malicious as an hyper-geometric distribution, i.e., 
\begin{equation}
P_{mn}(N,N_m,n,x)=\frac{\binom{N_m}{x} \binom{N-N_m}{n-x}}{\binom{N}{n}},
\label{eq:p_mn}
\end{equation}
\noindent

We can thus compute the probability that the packet recombined by the reference node during the $i$-th round is polluted as a weighted sum of\eqref{eq:recombined-polluted}, where the weights are the probabilities that $x$ out of $n$ uploaders are malicious, i.e.,
\begin{equation}
P_{gp}(i,N,N_m,n)=\sum_{x=1}^{n} P_{mn}(N,N_m,n,x) P_{rp}(i,x).
\nonumber
\end{equation}

Therefore, the probability that the reference node does not draw for recombination one of the polluted packets in its input buffer during any of the $\lceil \frac{k}{n} \rceil + 1$ rounds required to recover the generation is equal to
\begin{equation}
P_{fclean}(k,N,N_m,n)=\prod_{i=1}^{\lceil \frac{k}{n} \rceil + 1} 1 - P_{gp}(i,N,N_m,n).
\label{eq:pfclean}
\end{equation}

Finally, the probability that the reference node is able to recover a generation whose payload is not polluted is equal to
\begin{equation}
P_{rclean}(k,N,N_m,n)=\sum_{x=0}^{n} P_{mn}(N,N_m,n,x) P_p(\lceil \frac{k}{n} \rceil + 1,x,0).
\label{eq:prclean}
\end{equation}

\begin{figure}
  \begin{center}\rotatebox{270} {
	\includegraphics[width=0.5\columnwidth]{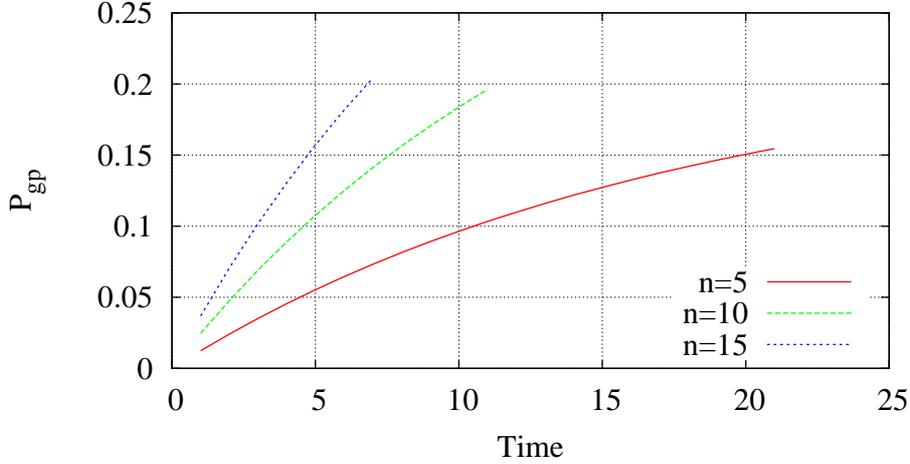}
	}
  \end{center}
  \caption{Probability that the packet transmitted by the reference node during the $i$-th round is polluted ($P_{gp}$) as a function of time ($k$=100).}
	\label{fig:pgp}
\end{figure}

The first observation we make is based on Figure~\ref{fig:pgp}, which shows the probability that the packet recombined by the reference node during the $i$-th round is polluted ($P_{gp}$) as a function of the time (i.e., the round index $i$) for a simple scenario like the one depicted in Figure~\ref{fig:model-scenario} with $N=1000$ nodes and $N_m=50$ malicious nodes, where each packet in the input buffer is recombined with probability $p_r=0.5$ and the probability that a malicious nodes transmits a polluted packet is equal to $p_{poll}=0.1$.
We observe that the reference node forwards a polluted packet to its downstream peers with a probability that increases with time: that is, packets forwarded later to downstream nodes are more likely to be polluted.
Therefore, downstream peers should draw for recombination each packet received by the reference node with a probability that is directly proportional with the age of the packet in the buffer (i.e., packets received earlier should be drawn for recombination with higher probability) rather than drawing each packet with identical probability $p_r$.

\begin{figure}
  \begin{center}\rotatebox{270} {
	\includegraphics[width=0.5\columnwidth]{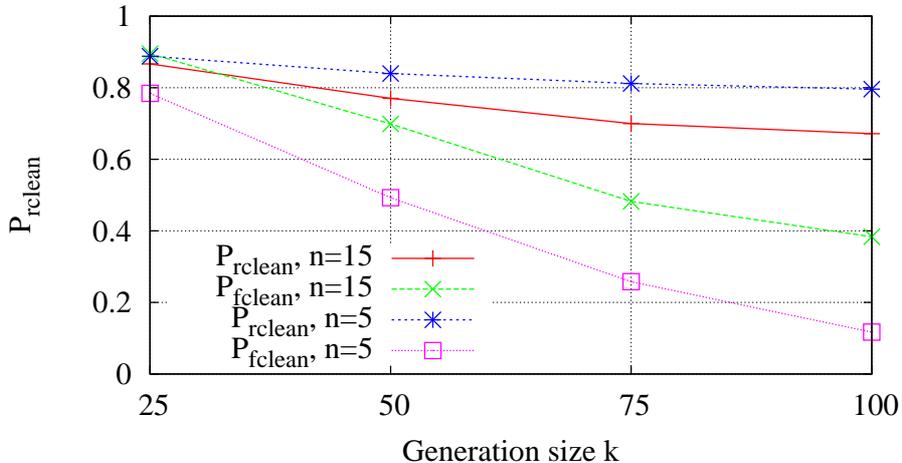}
	}
	\caption{Probability that a node forwards a clean packet $P_{fclean}$ and recovers a clean generation $P_{rclean}$ as a function of generation size $k$.}
	\label{fig:prfclean}
  \end{center}
\end{figure}

The second observation is that Equations \eqref{eq:pfclean} and \eqref{eq:prclean} both depend on one system parameter that can be controlled: the generation size $k$.
Figure~\ref{fig:prfclean} shows that small generations increase the probability to recover a clean generation and the probability of forwarding clean packets to downstream nodes.
Indeed, small generations reduce the overall number of rounds required to recover a generation (please remind that the number of rounds required by the reference node to recover the generation was assumed to be equal to $\lceil \frac{k}{n} \rceil + 1$ rounds).
The definition of $P_{fclean}$ is a product of probabilities, hence the lower the number of factors the higher the final results.
As for $P_{rclean}$, we note that $\forall x$, $P_p(\lceil \frac{k}{n} \rceil + 1,x,0)=(1-p_{poll})^{(\lceil \frac{k}{n} \rceil + 1)x}$ that is a decreasing function of the first argument that is equal to the overall number of rounds.
Short generations bring other advantages, such as reducing the computational complexity of recovering the coded payload \cite{bandcodestmm} and enabling low-delay communications by reducing the minimum required buffering time \cite{fiandrotti2012towards}, whereas a failure to timely recover a generation entails the loss of fewer video frames.
Note that, short generations also decrease the probability that received packets are innovative and may negatively affect the code overhead $\epsilon$.
However, as we experimentally demonstrate later on, small generations help reducing the pollution overhead to the point where the total network overhead is lower than for large generations.
Also, in Sec.~\ref{sec:proposed_architecture_recombination} we propose a simple heuristic that keeps the code overhead under control  by constraining the nodes to wait that a generation has reached a minimum rank before they start to recombine and relay the received packets. 

Concluding, the analysis of the results produced by our model suggest that:
\begin{itemize}  
\item packets received earlier by a node are less likely to be polluted than the following ones;
\item the probability that a generation can be correctly recovered increases as the generation size $k$ decreases;
\item the probability that a node transmits a polluted recombined packet decreases as the generation size $k$ decreases.
\end{itemize}
\noindent
Such findings represent the cornerstones of the pollution-resilient NC architecture described in the following section. 

%% file: architecture.tex
\section{Proposed Algorithms}
\label{sec:architecture}

In this section, we first describe a pollution detection scheme designed around On-the-Fly Gaussian elimination~\cite{bioglio2009fly} that allows a node to spot the presence of a polluted packet in the input buffer even before the generation is recovered.
Next, we present a packet recombination scheme that minimizes the likelihood that the packet transmitted by a node is polluted by exploiting
the knowledge unveiled by the model proposed in Sect.~\ref{sec:model}.

\subsection{Pollution Detection and Decoding}
\label{sec:detection-alg}

A basic feature of a pollution resilient NC P2P streaming application is the capability to detect that bogus data are being spread by unknown malicious peers.
Fortunately, we can exploit the NC decoding procedure, along with the fact that every node is likely to get some redundant (non innovative) packets from its neighbors, to obtain a pollution detection mechanism at generation-level. 
In other words, the algorithm described in the following paragraph allows every node to detect if a generation that is being decoded is under attack, albeit it cannot trace the pollution source.
We point out that that no ancillary data or infrastructure for verification are required and pollution detection is operated on the fly using only the received coded packets.
The algorithm operates in two stages, \emph{detection} and \emph{decoding}, that are detailed and described each in pseudo-code below.

The detection stage serves the purpose of revealing the presence of a polluted packet among those received by the node and detect whether received packets are innovative or not.
The detection stage is formalized as Algorithm~\ref{alg:ofg_stage_one} and it is executed every time a new packet $F_i= (y_i,g_i)$ is received by the node. In the following to avoid cluttering the notation we will drop the packet index using notation $F= (y,g)$ to refer a generic received packets.
Each time a node receives a packet, a copy of it is also stored in an input buffer for further recombination as described later on in this section.
The NC decoding process \cite{bioglio2009fly} can be represented as a solution to a system of $k$ linear equations $G X = Y$, where $G$ is a $k \times k$ upper-triangular matrix that stores (linear combinations of) the encoding vectors of the received packets,
$Y$ is the $k \times 1$ vector that stores the corresponding encoded payloads $y$ and $X$ is the $k \times 1$ vector that contains the symbols $x_i$ to recover, which are initially unknown.
In the following, we use the notation $G_i$ to indicate the $i$-th row of $G$ and we use the notation $G_{i,j}$ to indicate the element of $G$ at row $i$, column $j$.
When all the elements of $G_i$ and $Y_i$ are equal to zero, we say that the the $i$-th row of $G$ and the $i$-th element are empty and we write $G_i = \emptyset$.
Let $s$ be the index of the leading one of $g$, i.e. the first non-zero element of $g$ such that $g_i = 0 ~ \forall i < s$: the maximum number of iterations of the while cycle at line 2 of the algorithm is equal to $s$.
Depending on whether $G_s = \emptyset$, the algorithm operates as follows.
If $G_s$ is empty, $g$ is inserted in the $s$-th row of $G$, $y$ is inserted in the $s$-th position of $Y$ and the algorithm ends reporting an innovative packet was received (line 6).
Otherwise, a comparison between $G_s$ and $g$ is performed. 
If $g = G_s$, the received packet $P(g, y)$ and the pair $(G_s, Y_s)$ are expected to represent the same combination of the input symbols, thus the encoded payloads should match as well, i.e. it should be $y = Y_s$. This event occurs every time the packet being processed is linearly dependent on the ones received previously and it is likely to happen due to random coding, recombination and forwarding that imply the collection of $k' > k$ coded packets to complete decoding.
Therefore using the non innovative packet, a sanity check is performed comparing $y$ with $Y_s$: if they differ, then one or more packets received so far in the corresponding generation must be polluted and the algorithm returns reporting the presence of at least one polluted packet in the input buffer (line 9).
Otherwise, if payloads are identical, packet $F$ is likely to be correct but it is not helpful to recover the generation, so it is discarded and the algorithm returns reporting the received packet is not innovative (line 11).
If otherwise $g \neq G_s$, the algorithm performs a bitwise XOR between $g$ and $G_s$ and between $y$ and $Y_s$ (line 12): such XOR has the effect to set to zero the $s$-th element of the encoding vector, i.e. it sets $g_s = 0$, and the while cycle iterates unless any of the previously described termination is verified or $g_i = 0, \forall i$.  

\begin{algorithm}[h!]
  \caption{Pollution detection with Gaussian elimination}
  \label{alg:ofg_stage_one}
  \begin{algorithmic}[1]
  \STATE \textbf{receive}~$F=(y, g)$.
  \WHILE{true}
  \STATE $s \gets$ position of leading one of $g$.
    \IF {$G_s = \emptyset$}
      \STATE $G_s \gets g~~;~~ Y_s \gets y$
      \STATE \textbf{end}
    \ELSE
      \IF {$g = G_s$}
        \IF {$y \neq Y_s$}
          \STATE ~~~~\textbf{pollution detected; end};
        \ELSE
          \STATE ~~~~\textbf{useless packet; end}
        \ENDIF
      \ELSE
        \STATE $g \gets g \oplus G_s$; $y \gets y \oplus Y_s$
      \ENDIF
    \ENDIF
  \ENDWHILE
  \end{algorithmic}
\end{algorithm}
\noindent

The second stage, recovery, is executed when the rank of $G$ is equal to $k$, i.e. after $k$ linearly independent packets have been received.
Recovering the generation simply entails transforming the upper-triangular matrix $G$ as arranged during the detection stage to diagonal form by means of standard backward-substitution.
Algorithm~\ref{alg:ofg_stage_one} can be invoked each time a packet is received at the node, either before or after the generation has been decoded (due to the nature of push networks, nodes are likely to receive encoded packets also after they have recovered the generation).
In the following, we call \emph{early} packets received before the generation has been recovered; conversely, we call \emph{late} packets received afterwards.
If the algorithm is invoked to process early packets, we say that we have a case of \emph{early} pollution detection; otherwise, if the algorithm is invoked to process late packets, we talk about \emph{late} pollution detection.
In this latter case, late packets are exploited to double check whether any of the packets received so far was polluted.
Note that when Algorithm~\ref{alg:ofg_stage_one} returns a detected pollution flag, it is up to the node to decide how to exploit such information, for example during packet recombinations as described below.

\subsection{Packet Recombination at the Network Nodes}
\label{sec:proposed_architecture_recombination}

In this section we propose a packet recombination scheme that aims at reducing the probability that a packet forwarded by a node is polluted by exploiting the finding that packets received earlier are less likely to be polluted.
Let us assume that a node has received $r$ packets at the moment it is granted a transmission opportunity, and such packets are stored in a FIFO buffer as $\{F_1, \dots, F_i, \dots, F_r\}$, so that $F_i$ was received prior to packet $F_{i+1}$.
Each $i$-th packet is drawn for recombination according to packet recombination probability $p_r(i, \theta)$ that now we let depend on the packet index $i$; in particular, we propose to use the following truncated negative exponential density function

\begin{equation}
	\label{eqn:neg_exp}
p_r(i, \theta, \alpha) = \frac{i ^ {\alpha}}{\sum_{j=1}^{\theta} j ^ {\alpha}},
\end{equation}
\noindent
where $\alpha$ is the parameter of the exponential and $\theta$ is the cutoff parameter.

Now, the recombination vector $c = (c_1, \dots c_r)$, $c_i \in \{0,1\}$ defined in Sect.~\ref{sec:background}, is obtained by throwing $c_i$  as

\begin{equation*}
\begin{array}{c}
c_i = \left\{
\begin{array}{cl}
1 & \textrm{if~~~~$p_r(i, \theta, \alpha) < \rho$} \\
0 & \textrm{otherwise}
\end{array}
\right.
\end{array}
\end{equation*}
\noindent
where $\rho \in [0, 1]$ is drawn with uniform probability. 
The encoding vector of the recombined packet  is then computed as $g^r = \sum_{i=1}^{r} c_i g^i$, whereas the corresponding payload is computed as $y^r = \sum_{i=1}^{r} c_i y^i$ and finally packet $F^r=( y^r, g^r)$ can be forwarded to the neighbors.
Note that while the proposed scheme exploits the finding that packets received earlier are less likely to be polluted, we do not advocate that it globally minimizes the probability to transmit a polluted packet and we leave further improvements for future works.

Note that changing the recombination probability from a completely random one ($p_r=1/2$) to the time dependent function $p_r(i, \theta)$ may impair the coding overhead $\epsilon_c$ defined in Sect.~\ref{sec:background}.
In fact, drawing for recombination elder packets with higher probability limits the set of received packets that are recombined, decreasing the probability to create innovative packets.
To counter act this issue, we impose a minimum number of linearly independent packets $m_r$ that a node must have received for a generation before it is allowed to start forward linear combinations thereof.
At any time $m_r$ is equal to the rank of matrix $G$ in Algorithm \ref{alg:ofg_stage_one}  and allows us to put a lower bound on the cardinality of the set of packets used to generate novel recombinations.

%% file: torostream.tex
\section{The ToroStream P2P Protocol}
\label{sec:protocol}

In this section we overview the key aspects of ToroStream, a P2P protocol for live video streaming with NC that we use to evaluate our algorithms for pollution-resilient NC; a detailed description of the protocol can be found in our previous works~\cite{eusipco2012, bandcodestmm}, from which we borrow the terminology. Whereas in this work we use ToroStream to evaluate our proposed algorithms, in principle our algorithms can be applied to any NC-based P2P push or pull protocol.

\subsection{Topology Setup and Management}

Peer nodes are arranged into an unstructured, non-acyclic, mesh to minimize the topology management effort and increase the resilience to network failures.
A central tracker keeps track of all the nodes in the network: whenever a node wants to join the network, it contacts the tracker which replies to the node with a list of nodes already in the network drawn at random.
After a handshake, two nodes become neighbors and start to periodically exchange keepalive messages: if a node does not  receive keepalive messages from a neighbors for too long, the neighborhood relationship is terminated with an appropriate message.
The maximum size of the neighborhood of a node is upper bounded by $N_s$ so to maintain the network topology sparse and to minimize the related signaling and management overhead.
Also, periodically each node drops at random one or more nodes from its neighborhood to refresh the network topology.

\subsection{Signaling Protocol}

The server subdivides the video stream, which we assume encoded at constant bit rate $B_v$, into a sequence of independently recoverable generations of identical playout duration $C_t$ and approximately the same number $k$ of blocks of size $C_s$ each.
Every $C_t$ seconds, the server parses one generation of video from a video bitstream, subdivides the generation in $k$ blocks of symbols where the exact $k$ depends on the actual size of the video unit\footnote{In motion compensated hybrid video coding a simple way to recognize independently playable coding unit is always defined, and constitutes the so called group of pictures (GOP).} and distributes random linear combinations thereof to all its neighbors.
The generation currently distributed by the server is called the \emph{server position} in the following.
When a node joins the network, buffers $t_b$ seconds of video first, which correspond to $t_b/C_t$ generations, before playing out the generation with the earliest playout deadline in the stream.
The generation currently reproduced at the node is called here the node \emph{playback position}; generations encompassed between the server position (included) and the playout position of a node (excluded) form the \emph{decoding region} of the node.
Each node lets know its neighbors which generation within its own decoding region have already been recovered and which have not to its neighbors appending to all transmitted packets a vector of $t_b/C_t$ bits known as \emph{decoding map} which represents the decoding status of the generation within the node decoding region.

\subsection{Packet Scheduling and Pollution Avoidance Policy}

The server and the nodes distribute encoded packets with a random-push mechanism under a limited output bandwidth constraint as follows.
The server is allocated a maximum output bandwidth $B_s$: periodically, the server transmits a random linear combination of the blocks that compose the generation at the server position in the stream, where the transmission period is given by $B_s / C_t$.
The network nodes receive encoded packets which are processed for pollution detection and decoded as described in the the previous section and implemented as follows. 
Each time a node receives a packet, it stores a copy thereof in a separate input buffer for each generation in its decoding region.
Next, the packet is processed for pollution detection with Algorithm~\ref{alg:ofg_stage_one}: if the algorithm detects pollution, the corresponding generation is flagged as polluted.
The node keeps track of the status of each generation within its own decoding region with a vector of $t_b/C_t$ bits called \emph{pollution vector}, where each position of the vector is equal to one if any of the packets received for that generation was detected as polluted, 0 otherwise.
The pollution vector also drives the packet recombination mechanism of the network nodes as below.  
At each transmission opportunity, a node draws at random a node among its neighbors, checks the last decoding map received by that neighbor and performs a binary AND operation between the neighbor decoding map and its own pollution vector.
If all elements of the resulting vector are equal to 0, no generation is suitable for transmission either because at least one of the  packets in the corresponding input buffer is polluted at the node or because the neighbor has already recovered the generation. 
Otherwise, the node draws the generation suitable for recombination that is closer to the decoding deadline and recombines the received packet in the corresponding input buffer according to the algorithm described in the previous section and transmits the packet.

%% file: experiments.tex
\section{Experiments}
\label{sec:experiments}

In this section, we evaluate the pollution detection and packet recombination schemes proposed in Section~\ref{sec:architecture}
thorogh the random-push P2P protocol described in the previous section using a 64-cores server equipped with 128 GB of memory which hosts thousands of peers enbaling packet losses free experimentsing.
We consider a network of $N$ = 1000 nodes with $N_h$=980 honest nodes and $N_m$=20 malicious nodes, where the neighborhood of each node is restricted to $N_s$ = 25 nodes.
A 300 seconds test sequence encoded at $C_v$ = 500 kbit/s is distributed by a source node whose output bandwidth is equal to $B_s$ = 20 Mbit/s, whereas the output bandwidth of the peer nodes is constrained to $B_p$ = 750 kbit/s.
Peers implement the pollution detections scheme described in Section \ref{sec:architecture}: whenever a polluted packet is detected, the node stops transmitting packets for such generation to avoid further spreading the pollution.
All nodes enter the network  at the same time ($t=0$ s) and leave the network at the same time $t=300$ s.
Malicious nodes randomly alter the payload of each transmitted packet as described in Section \ref{sec:attack-model} and with probability $p_{poll}$ during the interval $[90, 210]$ s (\emph{attack interval}), whereas they behave as honest nodes, i.e. packets are altered with probability $p_{poll}$=0 for the rest of the experiment.
A generation is considered correctly recovered by a peer node if the node could timely recover the generation (i.e., if the node could receive at least $k$ independent packets) prior to its playout deadline and none of the received packets is actually polluted.
The quality of the video delivered to a node is measured in terms of Continuity Index (CI), which is defined as the fraction of generations that could be correctly recovered prior to the respective playout deadlines.

\subsection{Verifying the Pollution Model}

\begin{figure}[ht!]
	\begin{center}
	\subfigure{
		\rotatebox{270}{\includegraphics[width=0.35\columnwidth]{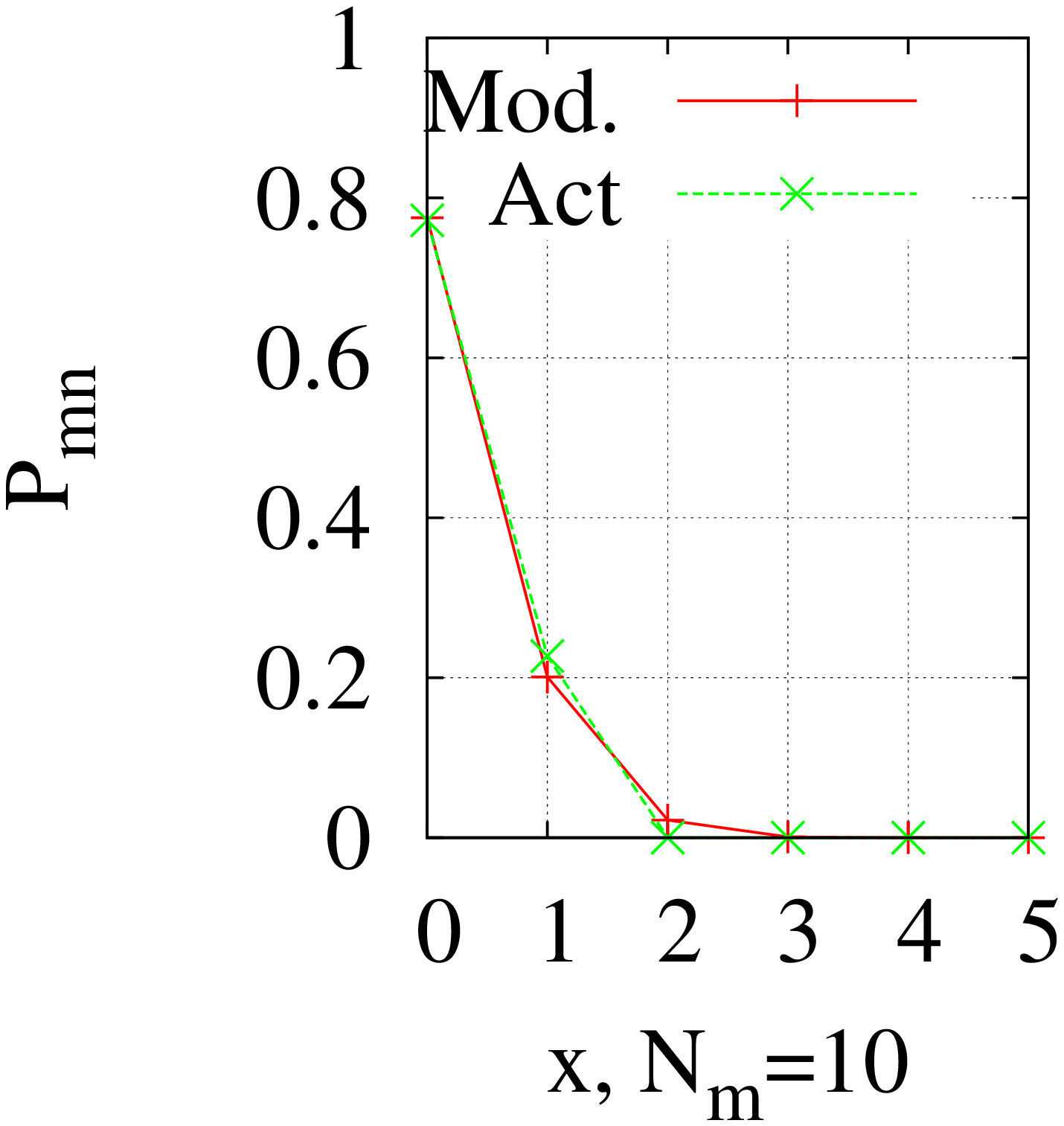}}
	}
	\subfigure{
		\rotatebox{270}{\includegraphics[width=0.35\columnwidth]{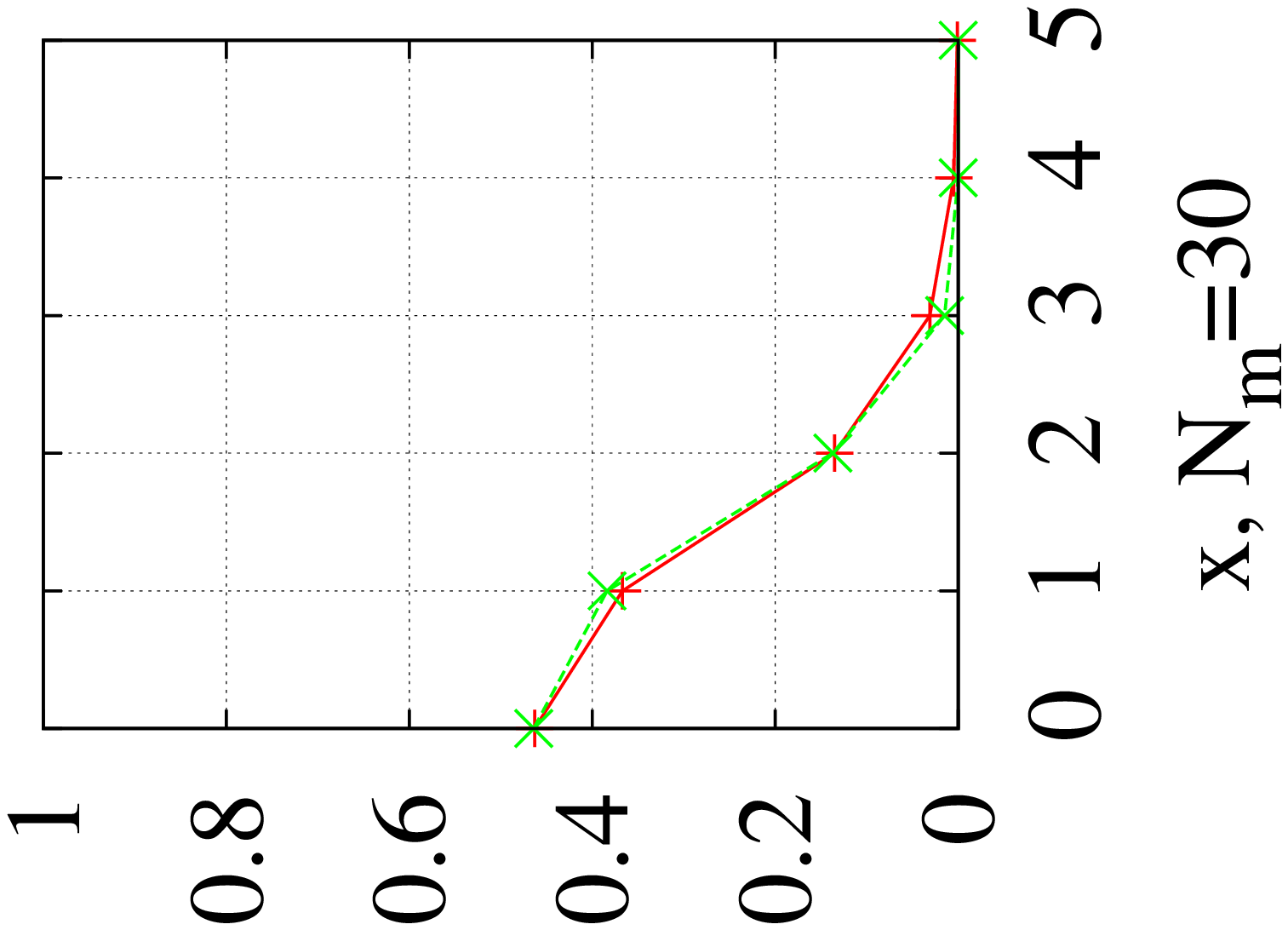}}
	}
	\subfigure{
		\rotatebox{270}{\includegraphics[width=0.35\columnwidth]{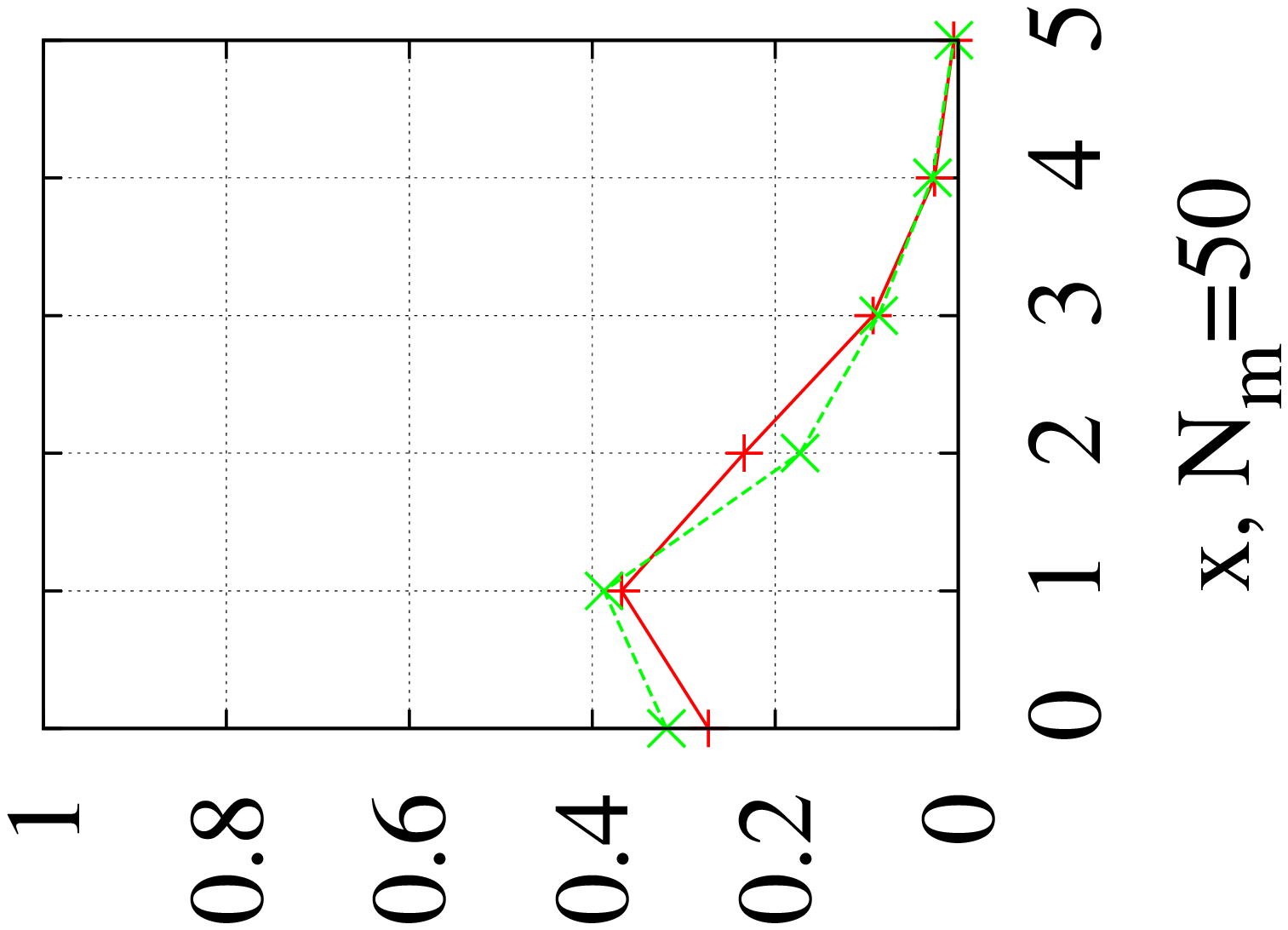}}
	}
	\end{center}
	\caption{Probability that  $x$ out of $n$ nodes are malicious for $N_m$ = 10 (left), 30 (center), 50 (right) nodes.}
	\label{fig:pmn}
\end{figure}

First, we verify the pollution model proposed in \ref{sec:attack-model} by sampling the actual distribution of malicious nodes among a node neighborhood.
We experiment in the above described scenario with $N$=1000 nodes, where each node has a neighborhood composed of $n$=25 other peers.    
Figure~\ref{fig:pmn} shows the expected and actual distribution of the probability that $x$ out of $n$ uploaders are malicious for different neighborhood sizes  $N_m \in [10, 30, 50]$ nodes are of the malicious type. We clearly see that probability that $x$ out of $n$ nodes are malicious follows an hyper-geometric distribution, as modeled in Eq. \ref{eq:p_mn}.

\vspace{-10pt}
\subsection{Effect of pollution attack on video quality}

\begin{figure}[h]
  \begin{center}
   \rotatebox{270} {
    \includegraphics[width=0.38\columnwidth] {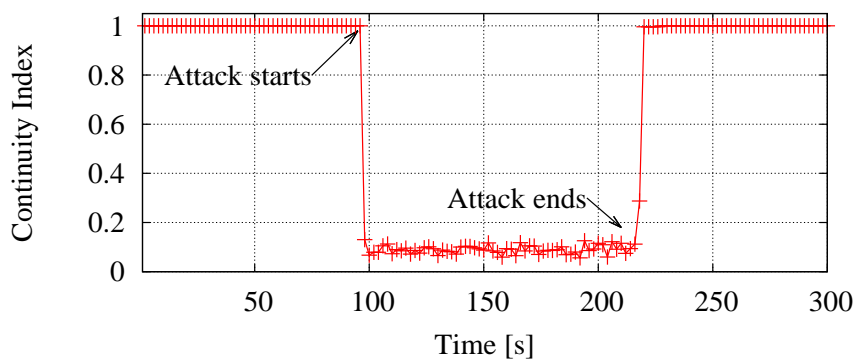}
   }
  \end{center}
  \caption{The video quality at the nodes drops in the 90$\sim$210 s interval due to the polluted packets transmitted by the malicious nodes.}
  \label{fig:ci_over_time}
\end{figure}

Then, we study the effect of a pollution attack for the reference NC architecture described in Section~\ref{sec:background}, where the peers recombine each received packet with probability $p_r=\frac{1}{2}$, malicious nodes alter the payload of transmitted packets with probability $p_{poll}$=0.01 during the attack interval and the video stream is subdivided in generations of $k$=50 blocks.
In this setup, the amount of packets purposely polluted by the malicious nodes amounts to about 0.02\% of the packets exchanged in the network.
Figure~\ref{fig:ci_over_time} shows the CI over time (each point in the graph corresponds to one generation).
During the time interval $[0, 90)$ no polluted packets are injected in the network by the malicious nodes and so the CI is equal to 1, i.e. all nodes decode the video without interruptions.
At time $t$=90 s, the 20 malicious nodes start injecting polluted packets for the following 120 seconds: during this interval, the CI drops from 1 to about 0.1.
Finally, at time $t$=210 s, malicious nodes cease transmitting polluted packets and the average CI rises again to 1 for the remaining 90 seconds of the experiment.
The CI averaged over the whole streaming session is equal to 0.628, whereas the average CI during the attack interval is equal to 0.111, i.e. about 9 generations out of ten cannot be correctly recovered due to the pollution attack.
A few malicious nodes are able to completely disrupt the communication by randomly altering less than 1\% of the overall network traffic, showing the need for countermeasures to pollution attacks.

\vspace{-5pt}
\subsection{Effect of pollution detection scheme}

\begin{figure}[h!]
  \begin{center}
   \rotatebox{270} {
    \includegraphics[width=0.52\columnwidth] {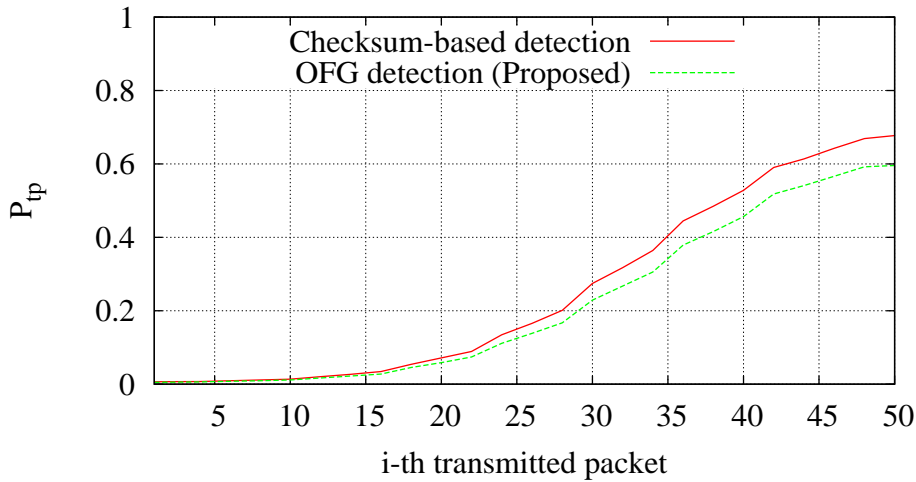}
   }
  \end{center}
  \caption{Probability $P_{tp}$ that the $i$-th packet transmitted by a node is polluted for different pollution detection strategies.}
  \label{fig:p_pkt_polluted_k_25_reference}
\end{figure}

First, we explore the effect of the pollution detection scheme on the probability that a honest node transmits a polluted packet further spreading the pollution in the network.
For this experiment, we consider the same reference packet recombination scheme as in the previous experiment and two different schemes for pollution detection.
The first scheme, \emph{OFG}, is our scheme described in Section~\ref{sec:detection-alg} where we exploit the OFG algorithm to verify if received packets are polluted even before the generation has not been recovered yet.
The second scheme, \emph{Checksum}, is an ideal strategy where the node recovers a generation, computes a checksum thereof and compares it with a reference checksum stored on a trusted server with unlimited bandwidth and zero latency.
Whenever pollution is detected for one generation, the nodes drop all received packets and stop relaying packets for that generation.
\\
Figure~\ref{fig:p_pkt_polluted_k_25_reference} shows the probability $P_{tp}$ that the $i$-th packet transmitted by a node is polluted.
The checksum scheme guarantees that a pollution attack is always detected at the moment a generation is recovered; however nodes must first recover the generation and only afterwards stop relaying polluted packets.
Conversely, our scheme provides no guarantee that a pollution attack is detected, however it can potentially detect pollution attacks and stop relying polluted packets earlier on.
Therefore, our OFG-based pollution detection is more effective that a checksum server-based reference in reducing the probability to relay polluted packets, plus nodes do not need to rely on a centralized checksum server with all the related issues.
\\
Moreover, we see that $P_{tp}$ is not constant, instead it grows over time with $i$ as predicted by our model and as shown in Figure~\ref{fig:pgp}, proving the qualitative correctness of the findings yield by our time-slotted model.
\\
This experiment shows that our OFG-based pollution detection scheme reduces the probability that an honest node relays a polluted packet, thus in all following experiments the nodes always implement our OFG-based pollution detection strategy.

\subsection{Effect of packet recombination strategy}

\begin{figure}[h!]
  \begin{center}
   \rotatebox{270} {
    \includegraphics[width=0.50\columnwidth] {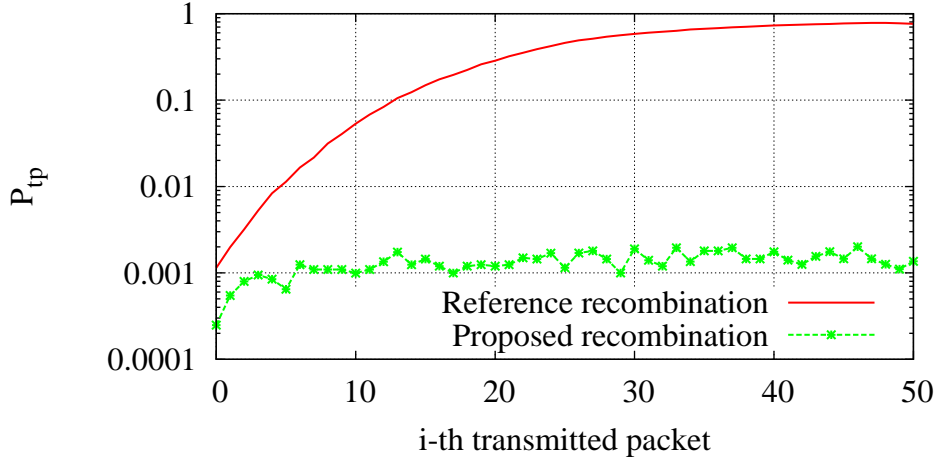}
   }
  \end{center}
  \caption{Probability $P_{tp}$ that the $i$-th packet transmitted by a node is polluted for different packet recombination strategies.}
  \label{fig:p_poll_over_n_k_50}
\end{figure}

Next, we study the probability that a node receives a polluted packet as a function of the packet recombination strategy at the nodes.
The first recombination scheme we consider is the same \emph{Reference} strategy used in previous experiments.
The second scheme, \emph{Proposed}, is our recombination scheme described  in Section \ref{sec:architecture}, where each $i$-th packet in the input buffer is drawn for recombination with a probability $p_r(i,\theta)$ that increases with the packet position $i$ in the buffer, i.e. with its age, as in Equation \ref{eqn:neg_exp} (in our experiments, we set $\alpha = 1$).
Unless stated in the following we use the proposed recombination algorithm with $m_r=1$, i.e. we do not put a constraint on the rank of the decoding matrix $G$.
Figure~\ref{fig:p_poll_over_n_k_50} shows the probability $P_{tp}$ that the $i$-th packet transmitted by a node is polluted.
With the reference packet recombination strategy, the probability that a node transmits a polluted packet quickly soars to about 0.8, i.e. almost 80\% of the packets in the network are polluted by the time the generation is recovered. 
Note that malicious nodes alter only about 0.02\% of the overall number of packets transmitted in the network, that is the reference strategy is responsible for an increase in the pollution rate of about 3 orders of magnitude.
Conversely, our proposed recombination scheme enables a $P_{tp}$ (about 0.12\%) which is two orders of magnitude lower than the reference scheme as packets received earlier, which are less likely to be polluted, are more likely to be drawn for recombination. 

\subsection{Effect of generation size}

\begin{figure}[ht!]
	\begin{center}
	\subfigure{
		\rotatebox{270}{\includegraphics[width=0.35\columnwidth]{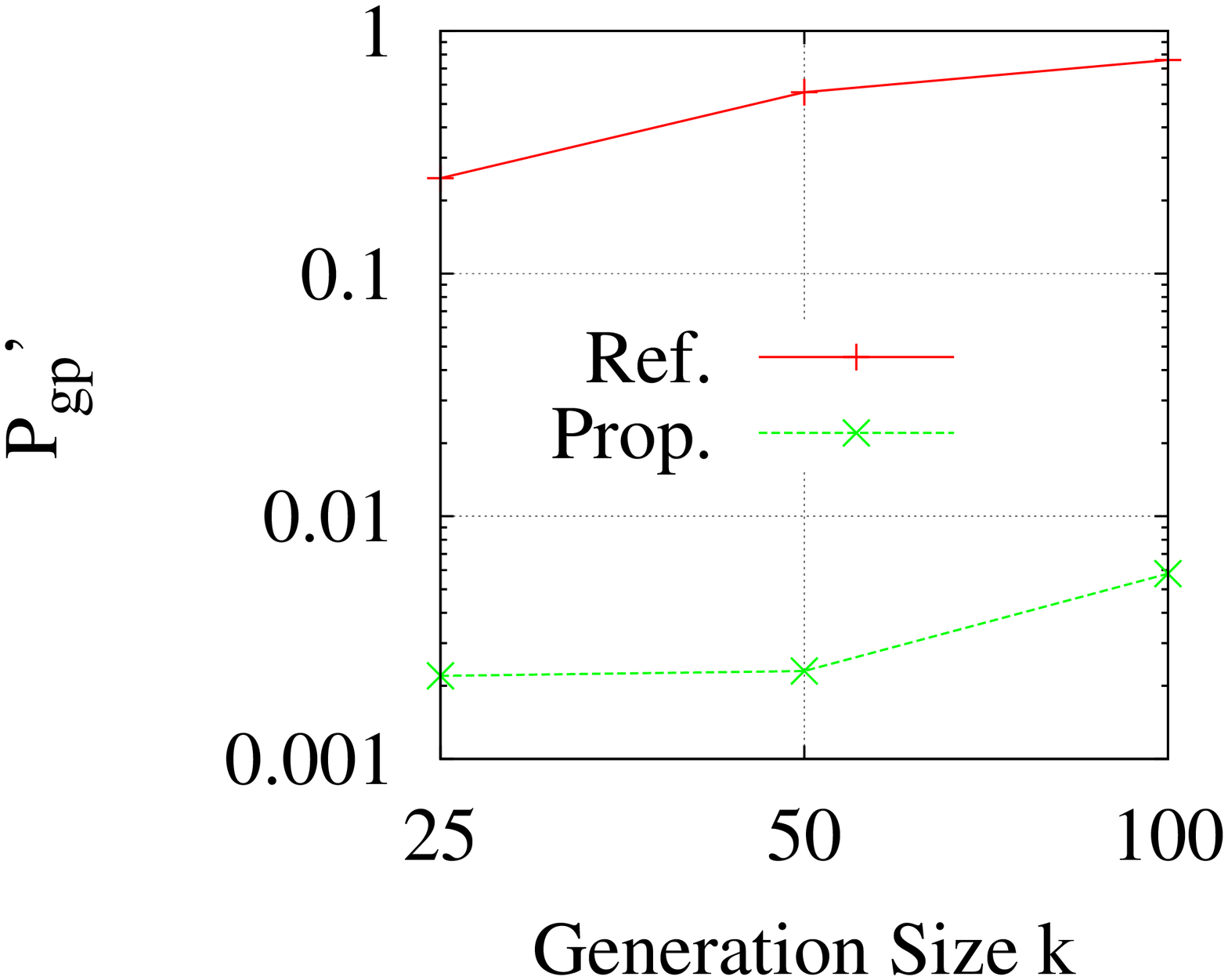}}
	}
	\subfigure{
		\rotatebox{270}{\includegraphics[width=0.35\columnwidth]{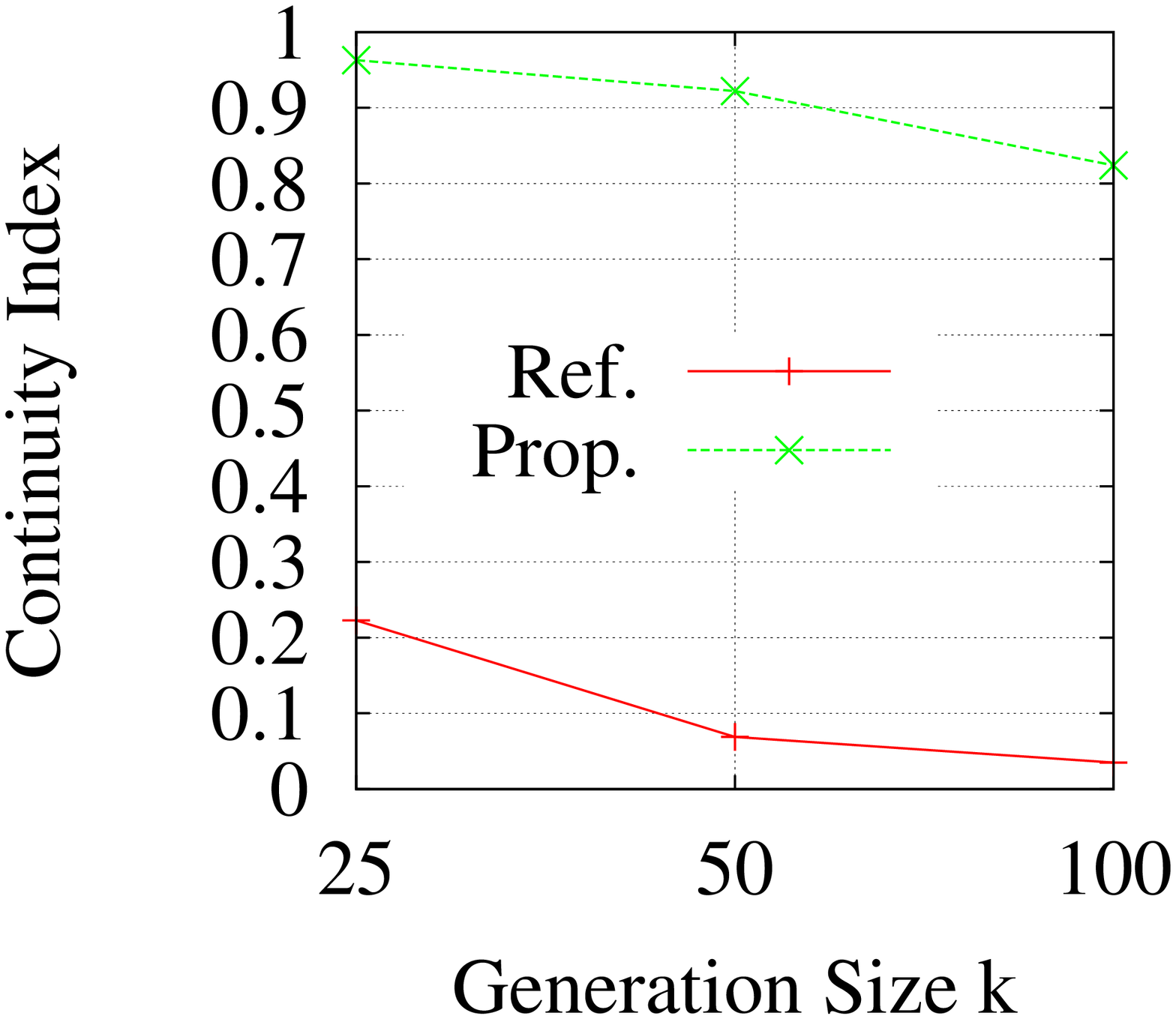}}
	}
	\end{center}
	\caption{Probability to transmit a polluted packet and corresponding video quality as a function of generation size $k$ for different packet recombination schemes.}
	\label{fig:ci_scheme_k}
\end{figure}

Next, in Figure~\ref{fig:ci_scheme_k} we evaluate the joint effect of the packet recombinations scheme and generation size $k$ on the probability that an honest node transmits a polluted packet $P_{tp}$ and on the CI and the relationship between the two.
Independently from the considered recombination algorithm, small $k$ yield lower $P_{tp}$ and thus higher CI as expected from Equation \ref{eq:prclean}.
However, just reducing $k$ is not sufficient to set off the pollution effects, and our packet recombination strategy is the key element in achieving near-optimal video quality. 
This experiments clearly demonstrates the relationship between the probability that a node transmits a polluted packet and the probability that the node is able to recover the generation.
In the following experiments, we experiment with the pollution attack model to assess the resilience of our scheme to an increased activity of the the malicious nodes.

\subsection{Effect of packet pollution probability}

\begin{figure}[h!]
  \begin{center}
   \rotatebox{270} {
    \includegraphics[width=0.54\columnwidth] {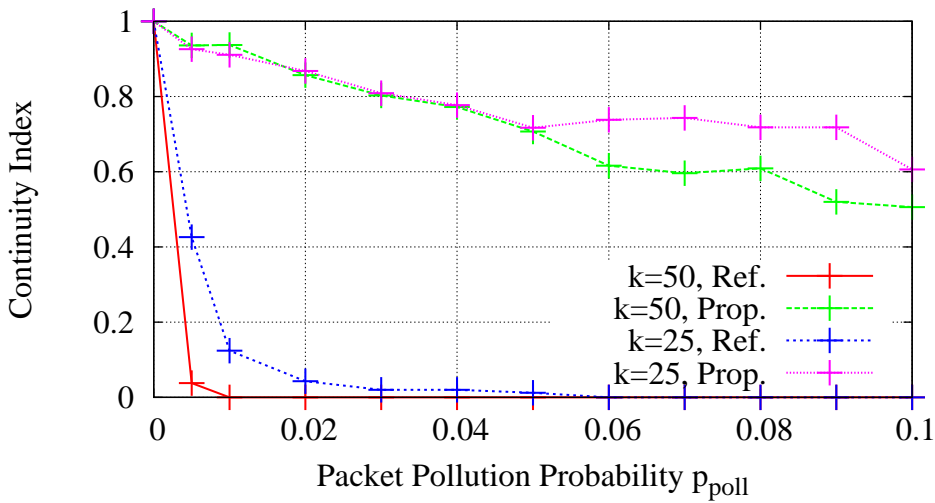}
   }
  \end{center}
  \caption{Video quality as a function of malicious nodes packet pollution probability for different packet recombination schemes and values of generation size $k$.}
  \label{fig:ci_p_poll_k}
\end{figure}

Figure~\ref{fig:ci_p_poll_k} shows the CI as a function of the probability $p_{poll}$ that a packet transmitted by a malicious node is  and for different packet recombination schemes and generation sizes $k \in \{25, 50\}$ (in all previous experiments we had $p_{poll}=0.01$).
As $p_{poll}$ increases, the CI drops to zero for the reference scheme, independently from $k$ (the larger $k$, the sharper the drop however).
Conversely, with our recombination scheme the video quality degrades gracefully despite a tenfold increase in the number of polluted packets transmitted to the network by the malicious nodes.
As expected, best video  quality is achieved when the proposed scheme is paired with smaller generations, albeit the largest contribution to pollution resilience is given by the our packet recombination algorithm.

\subsection{Effect of number of malicious nodes}

\begin{figure}[h!]
  \begin{center}
   \rotatebox{270} {
    \includegraphics[width=0.54\columnwidth] {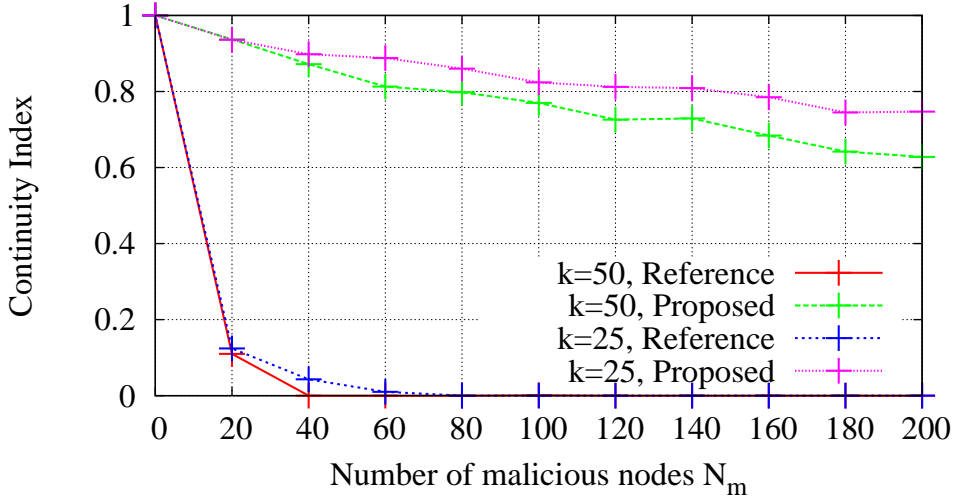}
   }
  \end{center}
  \caption{Video quality as a function of the number of malicious nodes in the network for different packet recombination schemes and values of generation size $k$.}
  \label{fig:ci_np_k}
\end{figure}

In Figure~\ref{fig:ci_np_k}, we investigate the relationship between video quality and number of malicious nodes $N_m$ present in the network.
As $N_m$ increases, the video quality drops to zero with the reference scheme, and reducing the generation size from $k=50$ to $k=25$  only marginally improves the performance.
Conversely, our proposed scheme allows a graceful degradation of the video quality as the number of malicious nodes in the network increases; moreover, small generations further improve the video quality.
Since the previous experiments confirm that the proposed recombination scheme paired with small generations yields best video quality, in the following we mainly focus on such combination of experimental parameters.

\subsection{Video Quality vs. Network Overhead Tradeoff}

\begin{figure}[ht!]
	\begin{center}
	\subfigure{
		\rotatebox{270}{\includegraphics[width=0.34\columnwidth]{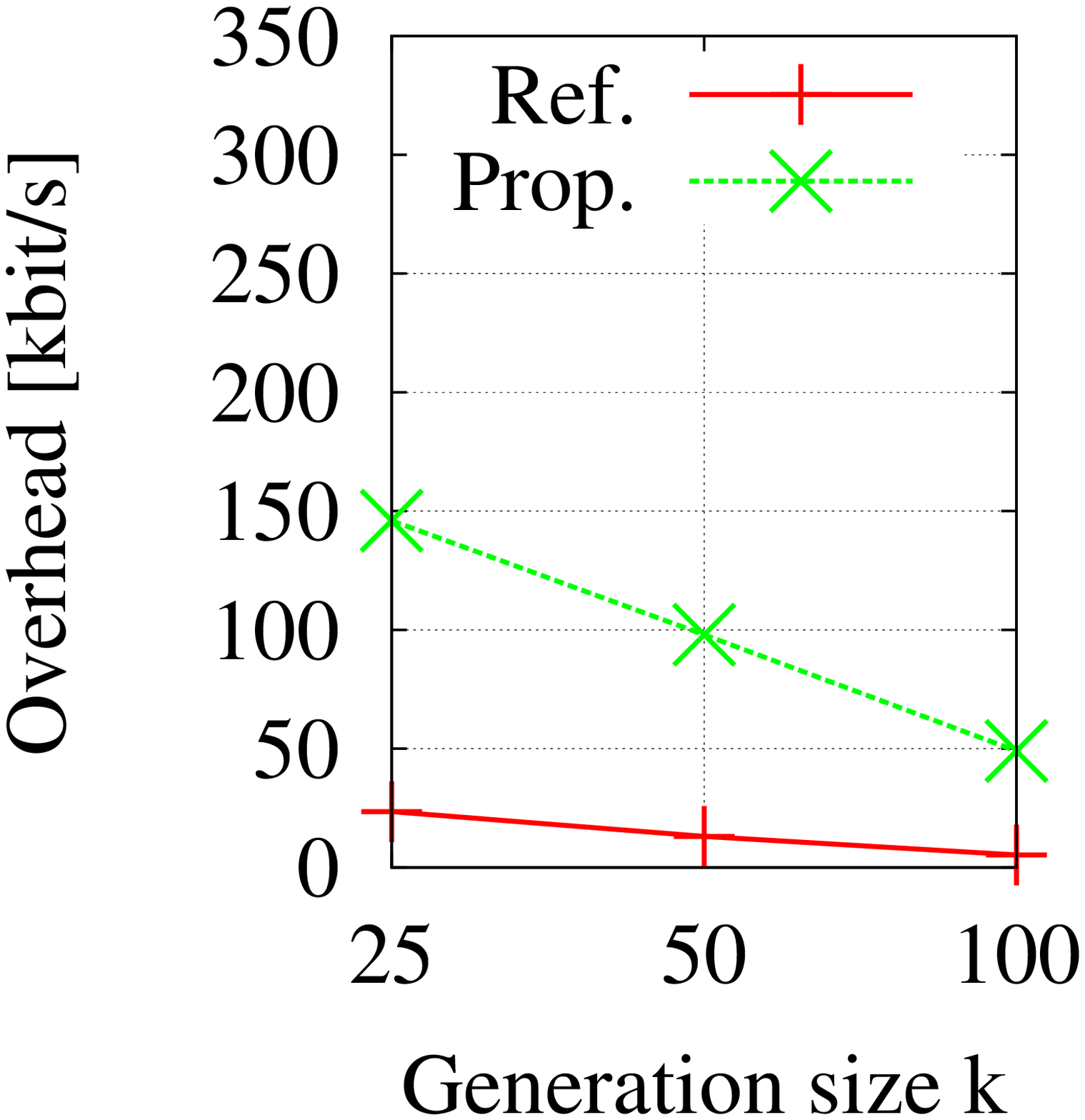}}
	}
	\subfigure{
		\rotatebox{270}{\includegraphics[width=0.34\columnwidth]{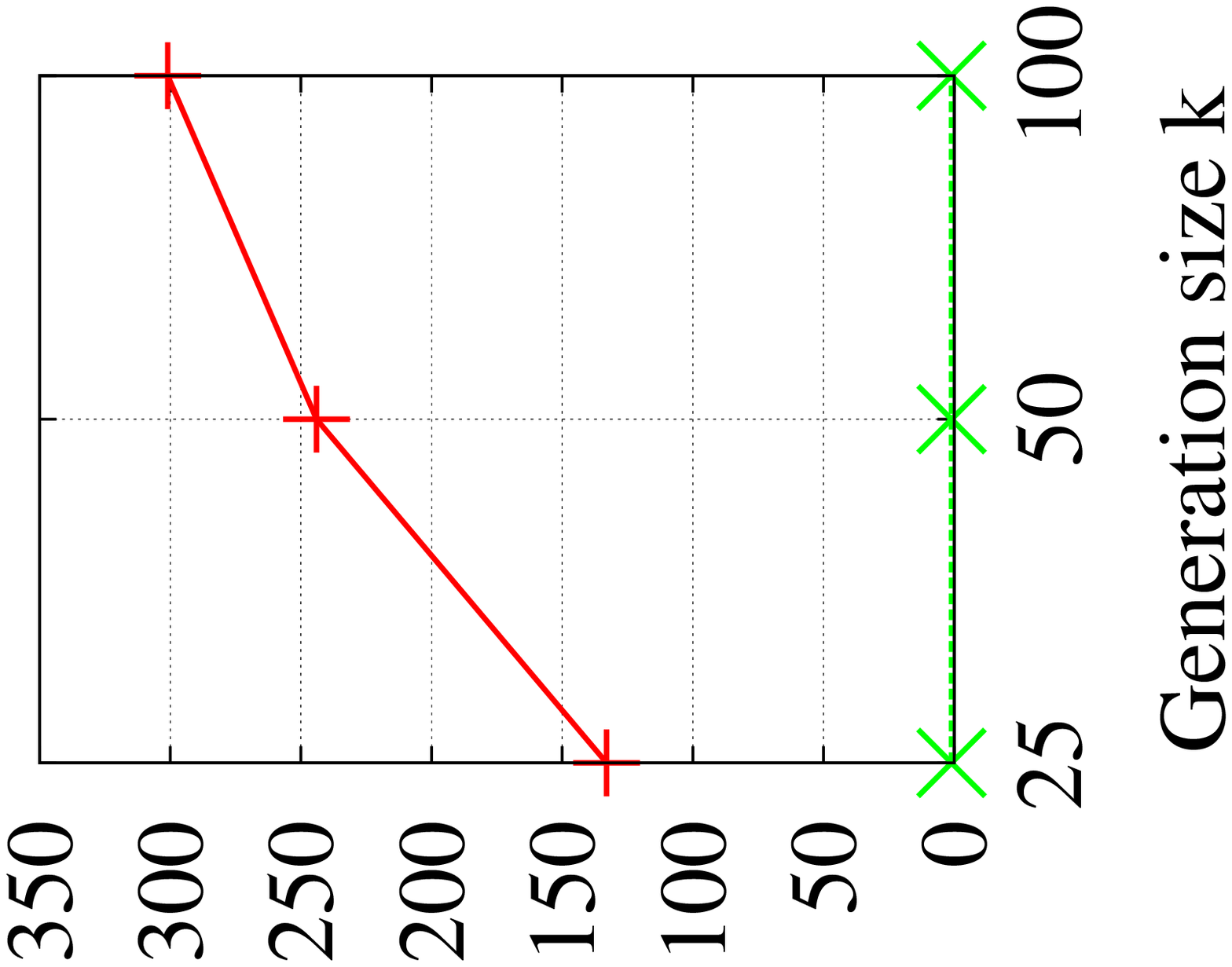}}
	}
	\subfigure{
		\rotatebox{270}{\includegraphics[width=0.34\columnwidth]{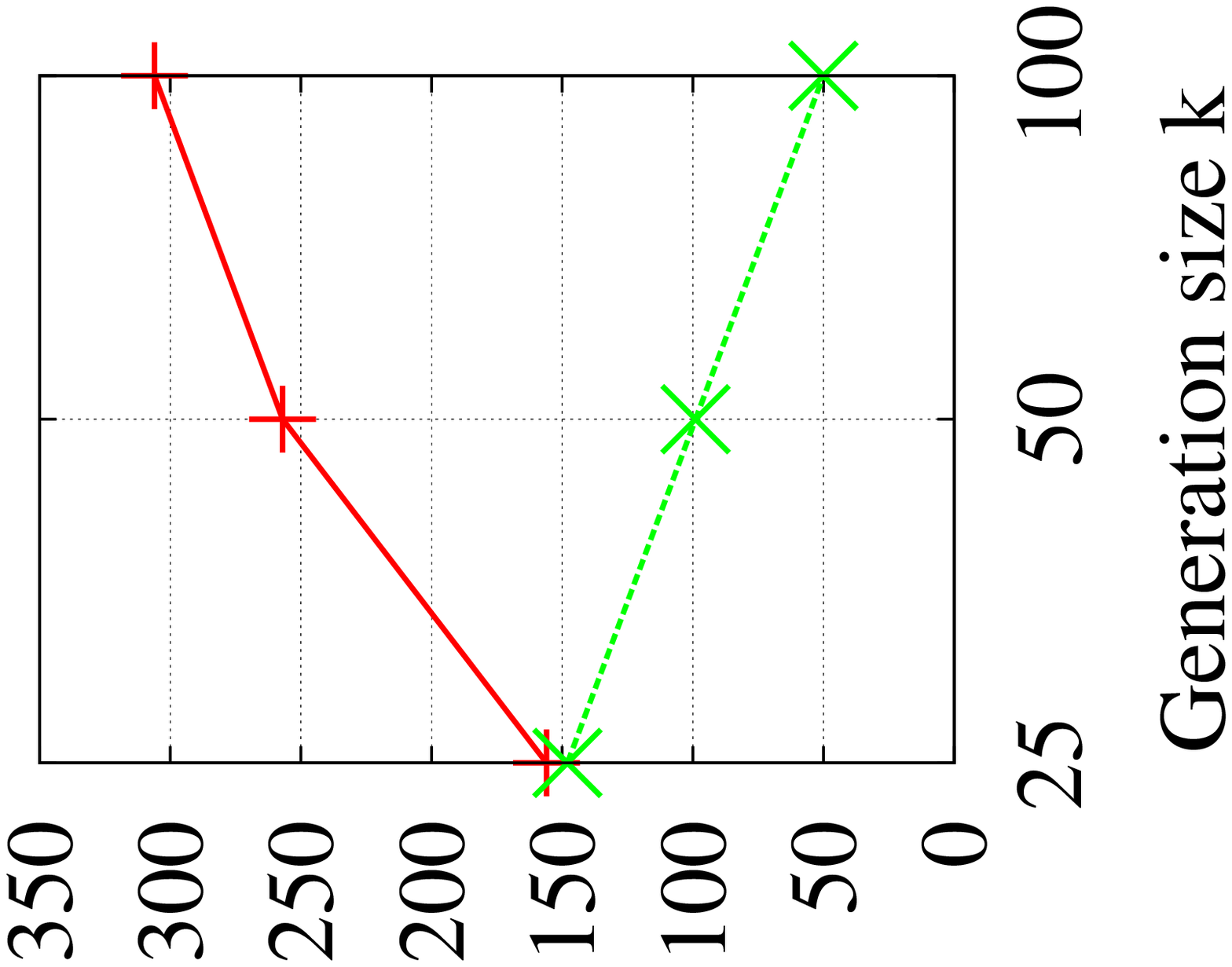}}
	}
	\end{center}
	\caption{\emph{Code} $\epsilon_c$ (left), \emph{pollution} $\epsilon_p$ (center) and \emph{total} $\epsilon_c$+$\epsilon_p$(right) overhead for the reference and proposed recombination strategy as a function of the generation size $k$. Proposed strategy reduces total overhead for any $k$.}
	\label{fig:strategy_k_overhead}
\end{figure}

Having shown that our proposed recombination scheme (with the help of small generations) sets off a pollution attack effect to the point where the video can be recovered almost seamless,  now we focus on the impact of the recombination scheme and generation size on the network overhead.
We recall that we define as code overhead $\epsilon_c$ the ratio of network bandwidth wasted transmitting non innovative packets; also the pollution overhead $\epsilon_p$ was defined as the ration of network bandwidth wasted transmitting polluted packets: the sum thereof is the total overhead, i.e. the overall ratio of wasted network bandwidth.
Figure~\ref{fig:strategy_k_overhead} shows, from left to right, the code, pollution and total overhead for three generation sizes $k$ and our packet recombination strategies plus the reference scheme.
As expected, short generations yield higher code overhead, regardless of the recombination scheme (left figure).
However, short generations help reducing the pollution overhead, plus our recombination scheme almost nullifies the pollution overhead as the central figure shows.
Therefore, as the right figure demonstrates, our proposed strategy yields a total overhead that is not higher than the corresponding overhead for the reference strategy even when generations are short, albeit it yields huge improvements in terms of video quality.

\begin{figure}[h!]
  \begin{center}
   \rotatebox{0} {
    \includegraphics[width=0.9\columnwidth] {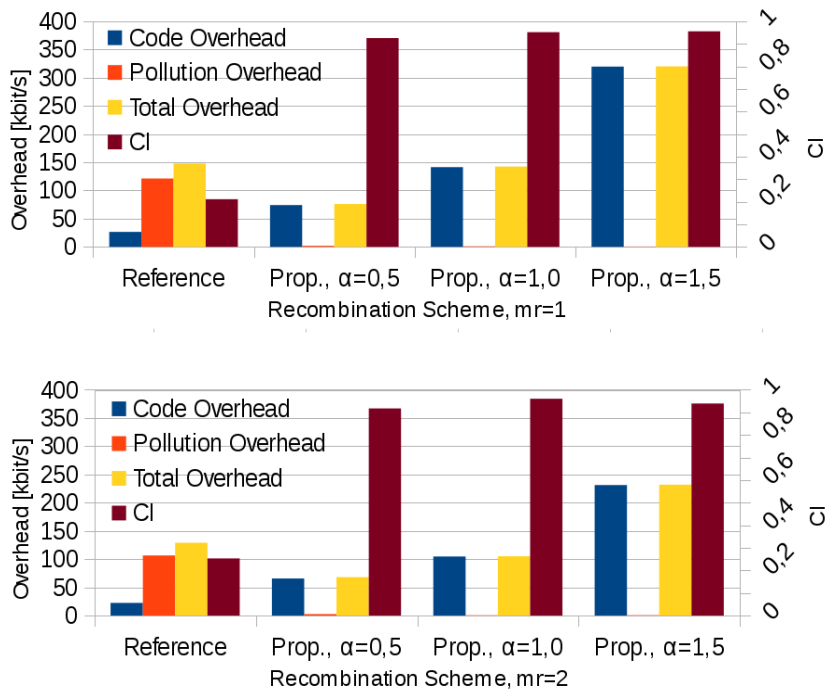}
   }
  \end{center}
  \caption{Tradeoff between video quality and network overhead for $m_r$=1 (top) and $m_r$=2 (bottom). For the proposed recombination strategies, three $\alpha$ values are considered (default in previous experiments is $\alpha = 1$).}
  \label{fig:ci_overhead_mr_alpha}
\end{figure}

Next, we investigate the video quality vs network overhead tradeoff as a function of two parameters of our packet recombination strategy.
In previous experiments, network nodes were allowed to start forwarding linear combinations of the received packets as soon as at least one packet was in the input buffer, i.e. $m_r=1$: we now experiment with $m_r$=2, i.e. nodes are allowed to transmit packets for a generation only if at least two linearly independent packets were received.
Moreover, in the previous experiments the $\alpha$ parameters in Eq. \ref{eqn:neg_exp} which controls the number of recombined packets for our proposed strategy was set to 1.0, i.e. we had $\alpha$=1.0: we now explore how the $\alpha$ parameter affects the performance of our scheme.
Figure~\ref{fig:ci_overhead_mr_alpha} shows the tradeoff between continuity index and total overhead, for the case $k$=25 and for the case $m_r$=1 (top) and $m_r$=2 (bottom).
\\
As previously seen, the reference strategy yields the largest pollution overhead, resulting in large total overhead and poor CI.
As $\alpha$ decreases from 1 to 0.5, more packets are recombined, thus the probability to recombine innovative packets increases and the code overhead drops while the CI is only marginally affected.
By comparing the top and bottom figures, we see that if nodes wait to receive a few independent packet before starting to relay, the code overhead drops independently form the considered recombination strategy.
In detail, this experiments shows that by controlling the $\alpha$ and $m_r$ parameters, we can further boost the performance of our strategy to achieve nearly optimal video quality and half the network overhead of the reference scheme.

%% file: stateoftheart.tex
\section{Related Works}
\label{sec:related}
As already pointed out in Sect.~\ref{sec:intro}, to the best of our knowledge the present paper is the first to face the P2P pollution problem from a novel point of view,  namely building a NC based P2P streaming application intrinsically resilient to the attack. Therefore, the goal here has been to mitigate as far as possible the effect of pollution, by leveraging on innovative use of the the NC decoder for pollution detection and by designing a novel pollution resistant recombination strategy.

Many research studies have proposed techniques to defend peer-to-peer streaming systems from pollution attacks following different strategies aiming at identifying malicious peers in order to remove them from the network. Clearly, such approaches are potentially the best solution to the the pollution problem; nonetheless,  malicious uploaders identification is very complex issue in random push NC based applications and the proposed techniques are usually limited by the number of polluters they can face or in terms of added computational complexity and/or communication overhead. In the following we provide a quick review of the related studies limited to the area of network coding.  

Several efforts have been devoted to devise on-the-fly verification techniques carried out by participants \cite{verify1,verify2,verify3,verify4,verify5,verify6,verify7}. 
These works are based on either  cryptographic or algebraic approaches. 
The major drawback of these elegant methods is the high computational costs for verification and the communication overhead due to pre-distribution of verification information.  Pre-distribution of verification keys is particularly critical in case of live streaming where novel data are being forwarded at a high rate. Error correction is another approach to deal with pollution attacks in network coding based peer-to-peer streaming \cite{correct1,correct2,correct3}; these methods introduce coding redundancy to allow receivers to correct errors but their effectiveness depends on the amount of corrupted information.

In \cite{Li_Lui} a fully distributed detection algorithm based on a stochastic approach is presented. The technique uses intersection operations to progressively isolate malicious peers in the set of neighbors of a peer.  The main drawback of the approach is that it works only under the (unrealistic) assumption that the neighbors remain the same and that each chunk is obtained by a randomly chosen subset thereof. 
In \cite{gaeta2014dip,gaeta2013identification}  malicious nodes identification is treated as an statistical inference problem relaying on control information termed check created by peers upon completing decoding of every chunk. Also in this case the additional communication and computational costs are needed. Moreover,  as in all statistical approaches the identification may fail leading to expungement of honest peers.

Finally, it is worth noticing that all previous approaches are exposed to the so called sybil attack, where malicious nodes try to escape identification by changing their identity at a pace higher than the identification mechanism rate. 

In the area of P2P file sharing, the injection of bogus data by untrusted peers has been traditionally  tackled using data authentication. In particular, a security hash,e.g. SHA1, can be computed  for each data block in order to recognize malicious modifications on the receiver side. Such approach must rely on a trusted infrastructure and protocol to distribute hashes to all peers in the network. Indeed, if the hashes distribution is not secure, malicious nodes can recompute and update the hash of a modified data to hide out.
Whilst being a viable approach for pull-based file sharing application such as BitTorrent, data hashing can not extended to video streaming  where real time computation and distribution of the verification data cannot be easily guaranteed. 

%% file: conclusions.tex
\section{Conclusions and future work}
\label{sec:conclusions}

In this paper we proposed simple countermeasures for mitigating the effects of pollution attacks in NC-based video streaming.
First, we model the diffusion of the polluted packets through the network due to the recombinations at the nodes: our analysis suggest that packets received earleier by a node are less likely to be polluted, while the chances that node recovers a clean generation decrease with the generation size.
On the basis of such findings, we devise a packet recombination scheme where packets are drawn with a probability that grows with the packet age in the nodes input queues.
Our experiments with P2P video streaming shows that, in a traditional NC context, a handful of malicious nodes can completely disrupt the video quality just by injecting a few polluted packets in the network.
Conversely, our proposed packet recombination algorithm, paired with small generations, makes the communication significantly robust to the activity of malicious nodes, which need to inject many more polluted packets in the network before the video quality strats to drop.
Our experiments also suggest that increased malicious nodes activity is the premise for devising effective mechanisms for detecting the malicious nodes and isolating them from the network, which we leave as future work.

%% file: ms.bbl
\begin{thebibliography}{10}

\bibitem{bioglio2009fly}
V.~Bioglio, M.~Grangetto, R.~Gaeta, and M.~Sereno.
\newblock {On the fly gaussian elimination for LT codes}.
\newblock {\em IEEE Communications Letters}, 13(12):953--955, 2009.

\bibitem{ross_07}
P.~Dhungel, X.~Hei, K.W. Ross, and N.~Saxena.
\newblock The pollution attack in {P2P} live video streaming: measurement
  results and defenses.
\newblock In {\em Proceedings of the 2007 workshop on Peer-to-peer streaming
  and IP-TV, P2P-TV '07}, pages 323--328, 2007.

\bibitem{bandcodestmm}
A.~Fiandrotti, V.~Bioglio, M.~Grangetto, R~Gaeta, and E.~Magli.
\newblock Band codes for energy-efficient network coding with application to
  {P2P} mobile streaming.
\newblock {\em IEEE Transactions on Multimedia}, 16(2):521 -- 532, February
  2014.

\bibitem{eusipco2012}
A.~Fiandrotti, A.~M. Sheikh, and E.~Magli.
\newblock Towards a {P2P} videoconferencing system based on low-delay network
  coding.
\newblock In {\em Proceedings of the 20th European Signal Processing Conference
  (EUSIPCO)}, pages 1529 --1533, 2012.

\bibitem{fiandrotti2012towards}
A.~Fiandrotti, A.~M. Sheikh, and E.~Magli.
\newblock Towards a p2p videoconferencing system based on low-delay network
  coding.
\newblock In {\em Signal Processing Conference (EUSIPCO), 2012 Proceedings of
  the 20th European}, pages 1529--1533. IEEE, 2012.

\bibitem{gaeta2013identification}
R.~Gaeta and M.~Grangetto.
\newblock Identification of malicious nodes in peer-to-peer streaming: A belief
  propagation-based technique.
\newblock {\em Parallel and Distributed Systems, IEEE Transactions on},
  24(10):1994--2003, 2013.

\bibitem{gaeta2014dip}
R.~Gaeta, M.~Grangetto, and L.~Bovio.
\newblock Dip: Distributed identification of polluters in p2p live streaming.
\newblock {\em ACM Transactions on Multimedia Computing, Communications, and
  Applications (TOMCCAP)}, 10(3):24, 2014.

\bibitem{verify2}
C.~Gkantsidis and P.~Rodriguez.
\newblock Cooperative security for network coding file distribution.
\newblock In {\em IEEE INFOCOM}, 2006.

\bibitem{p2p-nc-1}
M.~Grangetto, R.~Gaeta, and M.~Sereno.
\newblock Rateless codes network coding for simple and efficient {P2P} video
  streaming.
\newblock In {\em IEEE International Conference on Multimedia and Expo, 2009
  (ICME 2009).}

\bibitem{correct1}
T.~Ho, B.~Leong, R.~Koetter, M.~Medard, M.~Effros, and D.R. Karger.
\newblock Byzantine modification detection in multicast networks with random
  network coding.
\newblock {\em Information Theory, IEEE Transactions on}, 54(6):2798 --2803,
  june 2008.

\bibitem{huang2007pplive}
G~Huang.
\newblock {PPLive: A practical {P2P} live system with huge amount of users}.
\newblock In {\em Proceedings of the ACM SIGCOMM Workshop on Peer-to-Peer
  Streaming and IPTV Workshop}, pages 22 -- 28, 2007.

\bibitem{correct2}
S.~Jaggi, M.~Langberg, S.~Katti, T.~Ho, D.~Katabi, M.~Medard, and M.~Effros.
\newblock Resilient network coding in the presence of byzantine adversaries.
\newblock {\em Information Theory, IEEE Transactions on}, 54(6):2596 --2603,
  june 2008.

\bibitem{Jin}
X.~Jin and S.H.G. Chan.
\newblock Detecting malicious nodes in peer-to-peer streaming by peer-based
  monitoring.
\newblock {\em ACM Trans. Multimedia Comput. Commun. Appl.}, 6:9:1--9:18, March
  2010.

\bibitem{verify4}
D.C. Kamal, D.~Charles, K.~Jain, and K.~Lauter.
\newblock Signatures for network coding.
\newblock In {\em In Proceedings of the fortieth annual Conference on
  Information Sciences and Systems}, pages 3--14, 2006.

\bibitem{verify6}
E.~Kehdi and Baochun Li.
\newblock Null keys: Limiting malicious attacks via null space properties of
  network coding.
\newblock In {\em INFOCOM 2009, IEEE}.

\bibitem{correct3}
R.~Koetter and F.R. Kschischang.
\newblock Coding for errors and erasures in random network coding.
\newblock {\em Information Theory, IEEE Transactions on}, 54(8):3579 --3591,
  august 2008.

\bibitem{verify1}
M.~N. Krohn, M.~J. Freedman, and D.~Mazieres.
\newblock On-the-fly verification of rateless erasure codes for efficient
  content distribution.
\newblock {\em Security and Privacy, IEEE Symposium on}, 2004.

\bibitem{verify3}
Q.~Li, D.-M. Chiu, and J.C.S. Lui.
\newblock On the practical and security issues of batch content distribution
  via network coding.
\newblock In {\em Network Protocols, 2006. ICNP '06. Proceedings of the 2006
  14th IEEE International Conference on}, 2006.

\bibitem{Li_Lui}
Y.~Li and J.C.S. Lui.
\newblock Stochastic analysis of a randomized detection algorithm for pollution
  attack in {P2P} live streaming systems.
\newblock {\em Performance Evaluation}, 67(11):1273 -- 1288, 2010.

\bibitem{ross_05}
J.~Liang, R.~Kumar, Y.~Xi, and K.W. Ross.
\newblock Pollution in {P2P} file sharing systems.
\newblock In {\em IEEE INFOCOM 2005}, volume~2, pages 1174 -- 1185, march 2005.

\bibitem{p2p-nc-3}
S.~Mirshokraie and M.~Hefeeda.
\newblock Live peer-to-peer streaming with scalable video coding and networking
  coding.
\newblock In {\em Proceedings of the First Annual ACM SIGMM Conference on
  Multimedia Systems 2010, (MMSys '10)}, pages 123--132.

\bibitem{p2p-nc-2}
M.~Wang and B.~Li.
\newblock Network coding in live peer-to-peer streaming.
\newblock {\em IEEE Transactions on Multimedia}, 9(8):1554--1567, Dec 2007.

\bibitem{MIS}
Q.~Wang, L.~Vu, K.~Nahrstedt, and H.~Khurana.
\newblock {MIS}: Malicious nodes identification scheme in network-coding-based
  peer-to-peer streaming.
\newblock In {\em INFOCOM, 2010 Proceedings IEEE}, pages 1 --5, march 2010.

\bibitem{verify7}
Z.~Yu, Y.~Wei, B.~Ramkumar, and Y.~Guan.
\newblock An efficient scheme for securing xor network coding against pollution
  attacks.
\newblock In {\em INFOCOM 2009, IEEE}.

\bibitem{verify5}
Z.~Yu, Y.~Wei, B.~Ramkumar, and Y.~Guan.
\newblock An efficient signature-based scheme for securing network coding
  against pollution attacks.
\newblock In {\em INFOCOM 2008. The 27th Conference on Computer Communications.
  IEEE}, 2008.

\bibitem{zhang2005coolstreaming}
X.~Zhang, J.~Liu, B.~Li, and T.S.P. Yum.
\newblock {CoolStreaming/DONet: A data-driven overlay network for efficient
  live media streaming}.
\newblock In {\em proceedings of IEEE Infocom}, volume~3, pages 13--17.
  Citeseer, 2005.

\end{thebibliography}
